\documentstyle[aps,prb,psfig]{revtex}
\newcommand{\Eq}[1]{Eq.\ (\ref{#1})}
\newcommand{\Equation}[1]{Equation (\ref{#1})}
\newcommand{\Fig}[1]{Fig.\ \ref{#1}}
\newcommand{\Figure}[1]{Figure \ref{#1}}

\newcommand{\Subsect}[1]{Subsect.\ \ref{#1}}

\newcommand{\vect}[1]{\mbox{\boldmath$#1$}}
\newcommand{\bvec}[1]{\mbox{\boldmath $  #1 $}}
\newcommand{\eq}{\begin{equation}}
\newcommand{\en}{\end{equation}}
\newcommand{\eqn}{\begin{eqnarray}}
\newcommand{\enn}{\end{eqnarray}}
\newcommand{\CR}{\nonumber \\}
\newcommand{\pa}{\partial}

\newcommand{\D}{\delta}

\newcommand{\Om}{\Omega}

\newcommand{\bra}{\langle}
\newcommand{\ket}{\rangle}
\newcommand{\lm}{\lambda}

\newcommand{\cJ}{{\cal J}}
\newcommand{\cD}{{\cal D}}

\begin{document}

\title{Coupling between Electron Spin and Ferromagnetic
Clusters}
\author{N. Hamamoto, N. Onishi} 
\address{ Institute of Physics, Graduate School of Arts and Sciences
University of Tokyo, Komaba, Meguro-ku Tokyo 153-8902,Japan}
\author{G. Bertsch}
\address{ Department of Physics and Institute for Nuclear Theory, Box 351560
University of Washington, Seattle, WA 98195, USA }
\maketitle

\begin{abstract}
We analyze the adiabatic magnetization of ferromagnetic clusters in 
an intermediate coupling regime, where the anisotropic potential is
comparable to other energy scales. We find a non-monotonic behavior of
the magnetic susceptibility as a function of coupling with a peak.
Coriolis coupling effects are calculated for the first time;
they reduce the susceptibility somewhat.
\end{abstract}


\section{Introduction}                                                      

Magnetic properties of a wide variety of ferromagnetic clusters, e.g. 
Fe, Ni and Co in transition metals involving 3d electrons
\cite{fe,ni,reviewb,reviewa},
and Ru, Rh and Pd associated with 4d electrons \cite{gd,rh,reviewb}, are
studied using the Stern-Gerlach technique.
The observed deflection profile, caused by the interaction of
magnetization and  the gradient of field strength,
provides information about these intrinsic magnetic moment of the cluster
as well as its coupling to the other degrees of freedom in clusters.
The size of cluster is small enough to be regarded as a single domain system,
and the electrons involved form a single giant
magnetic moment; we will call it the super-electron spin in this paper.
It is quite interesting to see that, in such a low dimensional system,
some elements form ferromagnetic clusters, 
even though the bulk material is nonmagnetic. 
Also, the magnetization is strongly dependent on the number of
atoms in the clusters.
Therefore, it is important in the present stage of study to establish
a method of analysis for extracting the intrinsic magnetic moment
from the observed deflection profile, and this is the motivation 
for the present work.

Let us consider the experimental set-up.
First, the clusters formed by the laser evaporation,
are cooled by the helium gas, and then the clusters 
are expanded to form a molecular beam.
There is a problem in identifying the temperature experimentally.
In the present study we assume that the clusters are in thermal
equilibrium.
The density of clusters in beam is low, so that the clusters
may be assumed to be isolated beyond the equilibration zone.
Therefore each cluster stays in a certain quantum state in the beam.
Finally, the clusters enter into Stern-Gerlach magnet and are deflected
by the interaction of the gradient of magnetic field and the magnetic
polarization of super-spin induced by the magnetic field.
At the entrance of the magnet, strength of the field changes gradually in time,
and a time dependent interaction for the super-electron spin
causes a transion of the initial quantum state to other quantum states.
If the time dependence is sufficiently weak compared with
coupling of the spin to other modes,
the transition probability to other modes can be neglected.
This is called the adiabatic approximation.
In the present work, we calculate the profile and the magnetization
with this assumption.

If the electrons were completely decoupled
from other degrees of freedom such as rotational motion,
the deflection profile would be a flat horizontal distribution independent
of the field strength.
But this is not actually the case for the observed profiles.
A small coupling of the magnetic moment to internal coordinates of the cluster
gives rise to spin relaxation, making the profile different
from the flat distribution.
Hence it is important to make clear how various couplings produce 
observed deflection profiles.

The simplest model is superparamagnetism 
in which the population of the magnetic states are
proportional to a Boltzmann factor \cite{spara}.
In other words, the cluster rotation plays a role as a heat bath for
the super-electron spin in the magnetic field.
In practical analysis for extracting the giant magnetic moment,
the Langevin formula is widely employed. 
It assumes equilibrium with a thermal reservoir at a temperature
which is the same as the source of the cluster beam. 
However, it predicts a rather sharp deflection profiles which is 
quite different from the broad profile that are often observed.
Hence, the superparamagnetic model seems to be too simple for the analysis
of the experiments.

Another simple model is locked-spin model in which the super-electron
spin is frozen to the intrinsic orientation of the cluster, which
of course is free to rotate \cite{lockq,lockc,VB96}.
This model seems successful in reproducing the small peak observed near
zero deflection angle, which experimentalists call ``superparamagnetism''. 
But it is applicable only to Gd clusters and not general.
Furthermore, the model ignores the angular momentum of the super-electron
spin, which is known to give recoil effects in the Einstein-de Haas effect.

We propose an intermediate coupling model as a method to extract
the giant magnetic moment from the deflection profile \cite{VB96}.
This model covers the superparamagnetic and locked-moments models
as weak and strong coupling limits, respectively.
This paper is organized as follows: 
the intermediate model is explained and formulated in Sec. 2, 
numerical results of profiles and magnetization are presented
for various strength of coupling, strength of magnetic field and temperature
in Sec. 3.  
The conclusions and discussions are presented in Sec. 4.
\section{Intermediate Coupling Model}

In the present model it is supposed 
that all the electron spins are aligned in the same direction
through the exchange interaction.
Thus, the electron spins are in stretched coupling states
having a giant total spin $S=n_s N$, where $n_s$ is the number of
spins participating the magnetic moment an atom.
Since the magnetic moment is proportional to the total spin,
the cluster has a single giant magnetic moment expressed
as $\mu=g_{\rm s} S$, in terms of electronic gyromagnetic ratio $g_{\rm s}$.

Consider as a typical case the Fe$_{100}$ cluster
at temperature 100K .
This has a spin value of $S \simeq 100$ and a thermal rotational
angular momentum $R \simeq 600$ in units of $\hbar$.
These large values of $S$ and $R$ 
justify a classical treatment of the problem.
In fact, in a previous paper \cite{lockc} 
the classical treatment allowed the problem to be reduced a simple and 
transparent calculation.

this was examined for a simple case,
by utilizing adiabatic invariance which makes the calculation
easy and transparent.

However in a more general case, in which a fluctuation of
spin orientation with respect to the cluster is significant, 
we are unable to make a simple classical treatment.
Furthermore, in the classical treatment,
the spin angular momentum is ignored, and
thus angular momentum conservation is violated.
Since this is a doubtful approximation when $S$ is comparable to $R$,
we follow a fully quantum mechanical treatment.

Our Hamiltonian $H$ is expressed as sum of three terms
\begin{equation}
\label{t_hamiltonian}
H=H_{\rm rot}+H_{\rm coupl}+H_{\rm mag},
\end{equation}
where the terms are defined as follows.
The first term, $H_{\rm rot}$, stands for the rotational energy of the cluster
which is expressed as
\begin{equation}
\label{r_hamiltonian}
H_{\rm rot}=\sum_{i=1}^{3} \frac{\hat{R}_{i}^{2}}{2{\cal J}_i},
\end{equation}
where $\hat{R}_i$'s represent operators of three angular momentum components
referred to the body fixed frame, and ${\cal J}_i$'s express principal
moments of inertia. 
The vibrational modes are not taken into account, 
because the Debye-temperature,
(for instance, 500[K] for iron) is much higher than the source temperature.
In other words, the rotational motion is considered to work mainly
as heat bath in the spin relaxation.

The second term, $H_{\rm coupl}$, expresses a coupling potential
between the cluster and the super-electron spin,
which is originated from the crystal magnetic anisotropy energy
caused by molecular or crystal fields.
The simplest form of the energy is the uniaxial magnetic anisotropy
which, has already been examined in ref \cite{VB96}. 
Here we assume that the clusters have an internal cubic structure and 
the potential has cubic symmetry.
This is true for observation of the direction of easy magnetization
being [100],[010] and [001] for iron and nickel.
The anisotropy  constant is measured in the form
\begin{equation}
  E_{\alpha} = K_{1} (\alpha_{1}^{2}\alpha_{2}^{2} +
\alpha_{2}^{2}\alpha_{3}^{2} + \alpha_{3}^{2}\alpha_{1}^{2})
 + K_{2}\alpha_{1}^{2}\alpha_{2}^{2}\alpha_{3}^{2} + \cdots
\end{equation}
where
$\alpha_{i}^{2}$'s express  the direction cosines of super-electron spin.
The measured values in the bulk
are $K_{1} = 36$ [mK/atom] and $K_{2} = 13$ [mK/atom]
for iron, and $K_{1} =- 4$ [mK/atom] and $K_{2} = 0$ [mK/atom] for nickel.
We consider only the first term in the present calculation.
In order to formulate the anisotropic interaction,
let us start with the above classical picture for angular momentum variables,
in which $S'_i$'s commute.
\begin{eqnarray}
\label{coupling}
H_{\rm coupl} & = &  8uS^{\prime\,4}(\alpha_{1}^{2}\alpha_{2}^{2} +
\alpha_{2}^{2}\alpha_{3}^{2} + \alpha_{3}^{2}\alpha_{1}^{2}-\frac{1}{5})
 \nonumber \\
&=&-u\left\{ S^{\prime4}_1 +S^{\prime4}_2 +S^{\prime4}_3 -
6(S^{\prime2}_1 S^{\prime2}_2 + S^{\prime2}_2 S^{\prime2}_3
+S^{\prime2}_3 S^{\prime2}_1) \right. \nonumber \\
 & & \left. +\frac{3}{5}(S^{\prime2}_1 +S^{\prime2}_2
+S^{\prime2}_3)^2 \right\}
 \nonumber \\
 &=& -uS^{\prime4}\sqrt{\frac{32}{35}}\left\{C^{(4)}_4(\hat{\vect{S}})
+C^{(4)}_{-4}(\hat{\vect{S}})+\sqrt{\frac{14}{5}
}C^{(4)}_0(\hat{\vect{S}}) \right\},
\label{coupling_int}
\end{eqnarray}
Here $C_{\mu}^{(\lambda)}(\hat{\vect{S}})$ stands for the
spherical harmonic proportional to $Y_{\lambda \mu}(\hat{\vect{S}})$;
$S'_i$ is the spin component with respect to the three axes in the
body-fixed frame, $S_i$ denotes the spin component referred to
the laboratory system, and $u=2K_1/n_s^4 N^3$.
The electron spin prefers the direction of 4-fold axes of the cubic symmetry.
When the direction of the spin is along the 4-fold axes,
the value of $H_{\rm coupl}$ has minimum energy $-\frac{8}{5}uS^{\prime4}$.
And when the direction of the spin is along the eight axes of
$(\pm \frac{1}{\sqrt{3}}\pm \frac{1}{\sqrt{3}}\pm \frac{1}{\sqrt{3}})$
represented by body-fixed coordinate, the value of
$H_{\rm coupl}$ turns out to be energy maximum $\frac{16}{15}uS^{\prime4}$.
In quantum Hamiltonian, $\vect{S^\prime}$ are operators and we are interested
in their matrix elements.
The reduced matrix element
obtained through the Wigner-Eckart theorem is expressed by
\begin{equation}
\langle S_0 || \left[\hat{S} \right]^L || S_0 \rangle =
\left(\frac{1}{2}\right)^L \sqrt{\frac{(2S_0+L+1)!}{(2S_0-L)!}}.
\end{equation}
For example, for $L=1$, one obtains the familiar form 
$\sqrt{S_{0} (S_{0}+1)(2S_{0}+1)}$.
The interaction between the rotor and the super-electron spin
in \Eq{coupling} is given by
\begin{equation}
 \label{c_hamiltonian}
H_{\rm coupl}(\vect{S},\Omega)=\sum_{\kappa,m} \left[ S^{4}\right]^4_m
{\cal D}^{4\ast}_{m\kappa}(\Omega) A_{\kappa},
\end{equation}
where $\kappa$ takes the values of only 0 and $\pm4$
and $A_{\pm 4}=\sqrt{5/14}A_0=-u\sqrt{32/35}$.
The wave function of rotor can be expanded in terms of the
 \( {\cal D} \)-functions \cite{BM},
\begin{equation}
\phi_{R\mu}(\Omega)=\sum_{k}f_{Rk}{\cal D}^R_{\mu k}(\Omega).
\end{equation}
Since the Hamiltonian $H_{\rm coupl}$ is scalar, 
the total angular momentum $\vect{I}=\vect{R}+\vect{S}$ is still
a good quantum number for the first two terms of the Hamiltonians,
$H_{\rm rot}+H_{\rm coupl}$.
Accordingly, we select the base 
labeled by total angular momentum, and magnetic quantum number $I_z$. 
The basis is obtained by angular momentum coupling of 
the \( \cal D \)-function  \( {\cal D}^R_{\mu k}(\Omega) \) and
$ | S \sigma \rangle $ to $|I M \rangle$.
Therefore,  the total wave function of the rotor coupled with the
super-electron spin is expressed as
\begin{equation}
\label{base1}
\Psi_{\nu I M}=\sum_{Rk\sigma}
 \left\langle R \mu S \sigma | I M \right\rangle
 {\cal D}^R_{\mu k}(\Omega)f^{\nu I}_{R k}\left| S \sigma
\right\rangle, 
\end{equation}
where $\nu$ represents an index specifying states having the same $I M$.
The matrix element of the coupling term between the bases in
\Eq{base1} is estimated as
\begin{eqnarray}
\label{c_matrix}  
(H_{\rm coupl})^I_{Rk,R'k^{\prime}}& = & \sqrt{(2R+1)(2S+1)}
(-1)^{S+R'-I}
\nonumber 
\\ &  &
 A_{\kappa}^{\prime}\left\langle Rk4\kappa | R'k^{\prime} \right\rangle 
 W(RSR'S;I4), 
\end{eqnarray}
where 
 $\sqrt{2S_{0}+1}A_{\kappa}^{\prime}=
 A_{\kappa} \langle S_0 || \left[\hat{S}
 \right]^4 || S_0 \rangle $.
In our calculation we use potential strength parameter
$ u^{\prime}=A^{\prime}_{\pm 4}$.
Now, the wave function (\ref{base1}) and the energy in the source are
determined by diagonalizing $H_{\rm rot}+H_{\rm coupl}$.
Since there remains the $(2I+1)$-fold degeneracy in energy for $M$,
we may omit $M$ from the label for the energy 
$E_{\nu I}$.

The third term in \Eq{t_hamiltonian} represents
the interaction between an external magnetic field
and the moments of the super-electron spin
\begin{equation}
\label{m_hamiltonian}
H_{\rm mag}=-\vect{B} \cdot \vect{\mu} = -Bg_{\rm s}\hat{S_z},
\end{equation}
We choose the direction of the applied magnetic field 
as the axis of quantization ($z$-axis).
This interaction breaks rotational symmetry for the cluster, but
the magnetic quantum number $M$ defined above is still conserved due to
the choice of quantization axis.
From (\ref{base1}) and (\ref{m_hamiltonian}), the matrix elements between two states
are calculated as
\begin{equation}
\label{m_matrix}
 \left\langle\Psi_{\nu IM}|H_{\rm mag}|\Psi_{\nu'I'M}\right\rangle
  =-Bg_{\rm s}\sqrt{2I'+1}
 (-1)^{I'}\left\langle I'M10|IM \right\rangle h_{\nu I \nu' I'},
\end{equation}
where $M$ independent part $h_{\nu I \nu' I'}$ is expressed as
\begin{equation}
\label{m_matrix2}
h_{\nu I \nu' I'}
=\sum_{Rk}f^{\nu I \ast}_{Rk}f^{\nu^{\prime} I^{\prime}}_{Rk}
\sqrt{S(S+1)(2S+1)}(-1)^{R-S+1}W(ISI'S;R1).
\end{equation}

For simplicity, we set the moment of inertia around the intrinsic axes
to take the same value
$$ {\cal J}_1={\cal J}_2={\cal J}_3={\cal J} .$$
In this case, 
total wave functions are classified by an additional quantum number
$\pi_k=0,1,2,3$ ($k$ mod 4),
since $H_{\rm rot}$ is diagonal and $H_{\rm coupl}$ couples to only
the states having $k$-quantum numbers different by 4.
This point is different from uniaxial anisotropy,
in which $k$-quantum number is strictly conserved due to the axial symmetry.
The total wave functions in the magnetic field are expressed as
\begin{equation}
\left | \Phi_{\alpha \pi_k M}(B)\right\rangle = \sum_{\nu I}F^{\nu
I}_{\alpha \pi_k M}(B)\left | \Psi_{\nu I \pi_k M}\right\rangle,
\label{adiabatic_base}
\end{equation}
where $\alpha$ labels state in the same $\pi_k,M$.

To compare with the experiment, we calculate the average magnetization, 
which we regard as the ensemble average of $\langle S_z \rangle$.
This is the sum of the product of the expectation value of $S_z$ and
the occupation probability of each state. 
Now we proceed ahead with the important assumption 
that effect of the magnetic field is adiabatic
when the cluster passes through Stern-Gerlach magnet. 
As variation of the magnetic field is slow when the
cluster enters into and goes out from Stern-Gerlach magnet.
Firstly the clusters are retained in the source region. 
In this region, as the cluster ensemble is in thermal equilibrium,
the occupation probability of each quantum state is proportional to
the Boltzmann factor $\exp(-E_{\nu I \pi_k}/k_{\rm B} T)$.

The magnetization of each state in \Eq{adiabatic_base} is calculated as
\begin{equation}
  S_{z,\alpha(\nu I) \pi_k M}(B) =
  \langle \Phi_{\alpha(\nu I)\pi_k M} \mid
  \hat{S}_{z} 
  \mid \Phi_{\alpha(\nu I)\pi_k M} \rangle ,
\end{equation}
where $\alpha(\nu I)$ stands for the label of states connected adiabatically
with the states $(\nu I)$ defined in absence of the magnetic field. 
  
Under the adiabatic condition, any transition between energy levels
does not occur even if magnetic field is applied. 
Occupation probability of each quantum state is not altered during the flight.
Then the deflection profile is obtained by
\begin{equation}
  \label{profile}
  P(s,B,T) = \frac{1}{Z(T)} \sum_{\nu I \pi_k M} 
  \delta (s-S_{z,\alpha(\nu I) \pi_k M})
\exp \left( \frac{-E_{\nu I \pi_k}}{k_{\rm B} T} \right)
\end{equation}
where $Z(T)$ represents partition function given by
\begin{equation}
Z(T)=\sum_{\nu I \pi_k} (2I+1)
  \exp\left( \frac{-E_{\nu I \pi_k}}{k_{\rm B} T}\right). 
\end{equation}
The magnetization, the ensemble average
$\langle S_z \rangle_{\rm en.av.}$, is expressed by
\begin{equation}
\langle S_z \rangle_{\rm en.av.} =
    \frac{1}{Z(T)} \sum_{\nu I \pi_k  M}
\left\langle\Psi_{\alpha(\nu I)\pi_k M} \mid S_z
 \mid \Psi_{\alpha(\nu I) \pi_k M}\right\rangle
\exp \left( \frac{-E_{\nu I \pi_k }}{k_{\rm B} T} \right). 
\label{szenav}
\end{equation}

\section{Magnetic Susceptibility}
\label{Mag_Sus}

In almost all the Stern-Gerlach experiments with clusters 
the magnetic field is so weak that 
the magnetization shows linear dependence. 
Hence, it is important in the analysis of the experiment
to estimate the magnetic susceptibility.
Besides, the magnetic susceptibility for the superparamagnetism and
locked moment model has a well-known value; to examine the magnetic
susceptibility will give us how intermediate coupling model
describes the two extreme model.
Therefore, we calculate the magnetic susceptibility for intermediate
coupling model and discuss weak and strong coupling limit 
in the following subsections.

For the present, let us discuss the general expression of 
magnetic susceptibility.
Using Feynman theorem, we obtain the expectation value of $ S_z $ as
\begin{equation}
\left\langle \hat{S_z} \right\rangle
=-\left\langle \frac{\partial}{\partial (B g_{\rm s})}H_{\rm mag} 
\right\rangle\\
=-\frac{\partial}{\partial (B g_{\rm s})}\left\langle H_{\rm mag} 
\right\rangle. 
\label{par1}
\end{equation}
The magnetic susceptibility is expressed as
\eqn
\chi &=& \left. 
\frac{\pa}{\pa g_{\rm s} B}\left\langle \hat{S_z} \right\rangle_{\rm en.av.}
\right|_{B g_S=0}
=-\left. \frac{\pa^2}{\pa (B g_{\rm s})^2}\Delta E_{\rm en.av.}
 \right|_{B g_{\rm s}=0}\CR
&=& -\frac{2}{Z}\sum_{\nu I M}\sum_{\nu' I'}
\frac{|\bra \Psi_{\nu' I' M}|\hat{S_{\rm z}}|\Psi_{\nu I M}\ket |^2}
{E_{\nu I}-E_{\nu' I'}}
\label{par2}
\enn
where $\Delta E_{\rm en.av.}$ and $Z$
means the ensemble average of energy shift 
and partiation function, respectively.

We will deal with a certain limit of susceptibility (\ref{par2})
in the following two subsections.
Strong coupling limit will be
discussed in \Subsect{susceptSC}.
The weak coupling limit will be treated in \Subsect{susceptWC}. 
Susceptibility in intermediate coupling will be
calculated numerically in \Subsect{susceptIC}.

\subsection{Susceptibility for Strong Coupling}
\label{susceptSC}

We first consider the energy eigenvalue and the eigenstate of 
$H_{\rm rot}+H_{\rm coupl}$ in the strong coupling limit.
The angular momentum of the rotor $R$ is mixed,
while total angular momentum $I$ and the $z$-component 
of total angular momentum,
$M$ are conserved in the absence of the magnetic field.
We rewrite the Hamiltonian of the present model 
in terms of the total angular momentum, $I$
\eq
\frac{\hat{\vect{I}}^2+\hat{\vect{S'}}^2
-2\hat{\vect{I}} \cdot \hat{\vect{S'}}}{2\cJ}+H_{\rm coupl}(\hat{\vect{S'}}).
\label{I_repr}
\en
The third term of the numerator is the Coriolis term, which
couples the degrees of freedom of super-electron spin to one of the rotor.
The Coriolis term is not taken into account in the locked moment model,
because the super-electron spin is not included as a dynamical variable.
On the other hand, it must be noticed that the intermediate coupling model
includes the Coriolis term. As will be seen this term contributes
to the magnetic susceptibility.
Since $H_{\rm coupl}$ gives a dominant contribution to the energy eigenvalue 
of $H_{\rm rot}+H_{\rm coupl}$ in the strong coupling, 
we treat the Coriolis term as a perturbation.
The coupling Hamiltonian $H_{\rm coupl}(\hat{\vect{S'}})$ 
is written in terms of the intrinsic component only (see \Eq{coupling_int}).
Then it is advantageous to select a direct product
the eigenfunction of total angular momentum and its 
$z$-component and
eigenfunction of super-electron spin with respect to intrinsic frame;
$\cD^I_{MK}(\Om) \otimes \sum_{\sigma_k} g_{\sigma_k}|S \sigma_k \ket$
as the basis.
The energy eigenvalue of the Hamiltonian neglecting Coriolis term is
\eq
\frac{\hbar^2}{2\cJ}(I(I+1)+S(S+1))+u' E^{\rm A}_N,
\label{band}
\en
where  $u' E^{\rm A}_N$ is the energy eigenvalue of the coupling
$H_{\rm coupl}$. 
Without the Coriolis interaction,
the energy spectra is of rotational band with band head energy
$u'E^{\rm A}_N$.

Let us now discuss the ground band of the present model
in the strong coupling.
\Figure{u_energy} illustrates the numerically evaluated 
energy eigenvalues of $H_{\rm coupl}$ in the space $|S\sigma \ket$.
The eigenstate of $H_{\rm coupl}$ must belong to a certain irreducible
representation of the point group $O$.
The irreducible representations $A_1,E,T_1$ for $S=4n$,
$A_2,E,T_2$ for $S=4n+2$,
and irreducible representations $T_1,T_2$
for odd $S$ appear in the lowest energy region.
Considering the dimension of these representations, namely
1,1,2,3 and 3 for $A_1,A_2,E,T_1$ and $T_2$ respectively,
the bunch of the levels contain six states.
The six fold and the eight fold approximately degenerate states 
appear at the lowest and highest energies, respectively.
This fact implies that the direction of super-electron spin is
localized to the six (eight) directions corresponding 
to the potential minima (maxima)
in the lowest (the highest) bunching states for large $S$.
Therefore, these states are approximated by
\eqn
R^i | SS \ket,
\label{locked_GS}
\enn
where $R^i$ stands for the operator corresponding to 
the rotation from 3rd axis to $i$th direction of the potential minima.

For the present, we shall take into account the Coriolis term
in first-order perturbation theory.
According to first order perturbation theory of degenerate states,
we solve the secular equation
to determine the energy shift and perturbed states.
In this case, each level in the rotational band is 
$6(2I+1)^2$-fold degenerate:
The quantum number $M,K$ have the values $-I\le M ~ K \le I$
and the ground state of super-electron spin is 6-fold degenerate.
The dimension of the secular equation is decreased by a factor $(2I+1)$ 
because the motion under the total Hamiltonian conserves $M$.
As we discussed before, the direction of super-electron spin corresponds to
one of the potential minima for the ground state.
We choose the direction of quantization axis as
one of the potential minima.
This selection of quantization axis yields the Coriolis matrix element as
\eqn
\bvec{\hat{I}\cdot \hat{S}}=\hat{I_i} \hat{S_i}
+ (\hat{I_+^i} \hat{S_-^i} + \hat{I_-^i} \hat{S_+^i})
\label{modified_Coriolis}
\\
\hat{I_i}=R^i \hat{I_3} R^{i \dagger} ,\quad 
\hat{I_\pm^i} = R^i \hat{I_\pm} R^{i \dagger},\CR
\hat{S_i}=R^i \hat{S_3} R^{i \dagger} ,\quad 
\hat{S_\pm^i} = R^i \hat{S_\pm} R^{i \dagger} \nonumber.
\enn
Only the first term of \Eq{modified_Coriolis} is involved in the 
calculation of the matrix element of Coriolis term between
the parallel direction of super-electron spin.
The second terms do not contribute 
the Coriolis matrix element between parallel or antiparallel direction
but orthogonal direction.
Actually, we can estimate the matrix element of raising 
and lowering operators of super-electron spin
\begin{equation}
\bra SS | \hat{S_\pm} R'(\theta)| SS \ket =\cases{%
0,& \quad \mbox{for $\theta =0$}\cr
\propto \hbar \sqrt{2S}(1/2)^{S+1},& \quad \mbox{for $\theta =\frac{\pi}{2}$}\cr
0,& \quad \mbox{for $\theta =\pi$}\cr
}
\label{offdiagonal}
\end{equation}
where $R'(\theta)$ is the rotation through angle $\theta$ around 
the axis perpendicular to the 3 axis.
We find from \Eq{offdiagonal} that the second term of the Coriolis term
( \Eq{modified_Coriolis}) is vanishingly small for large super-electron spin.
This allows us to approximate the Coriolis term as
\eq
\hat{\bvec{I}}\cdot \hat{\bvec{S}} \simeq \hat{I_i} \hat{S_i} .
\label{app_Coriolis}
\en
We can solve the secular equation for approximated 
Coriolis term, \Eq{app_Coriolis}, for each degenerate rotational state.
The wave function results in the eigenstate of $I^i$;
\eq
|IMN\xi \ket= |IMKi \ket 
= R^i | SS \ket R^i \sqrt{\frac{2I+1}{8\pi^2}}\cD^I_{MK}(\Om)
\quad (N \le 6).
\label{WF_coriolis}
\en
where $\xi$ specifies the energy levels split by the Coriolis term.
For the ground band $(N \le 6)$, $\xi$ corresponds to $K$, and
$N$ is labeled by the index of the direction of super-electron spin $i$.
The first order energy shift is obtained as
\eqn
\bra I'MK'j | 2 \hat{\vect{I}} \cdot \hat{\vect{S}} | IMKi \ket 
=2\hbar^2 KS\delta_{K,K'}.
\label{ground_Coriolis}
\enn

We neglect the coupling between ground band and excited band because
the energy difference between these bands are large in the strong coupling.
Higher order perturbation would take into account this coupling and 
would give the deviation from the locked moment due to the finite strength
of coupling. But we do not discuss the deviation in this paper.

The energy eigenvalue of (\ref{I_repr}) is divided into three parts
\eq
E_{IMN\xi}=\frac{\hbar^2}{2\cJ}(I(I+1)+S(S+1))+u' E^{\rm A}_N
+\frac{\hbar^2}{2\cJ}E_{N\xi}^{\rm C}
\label{E_eigen}
\en
where $E^{\rm C}_{N\xi}$ is the energy shift by the perturbation 
of the Coriolis term and is equal to \Eq{ground_Coriolis} 
for the ground band ($N \le 6$)

The susceptibility is evaluated through \Eq{par2} as 
\eqn
\chi &=&
-\frac{2}{Z}\sum_{IMN\xi} \sum_{I'N'\xi'}
\frac{|\bra I'M N'\xi'|\hat{S_z}|IMN\xi \ket |^2}
{\frac{\hbar^2(I(I+1)-I'(I'+1))}{2\cJ}
-\frac{\hbar^2(E^{\rm C}_{N\xi}-E^{\rm C}_{N'\xi'})}{2\cJ}
-u'(E^{\rm A}_N - E^{\rm A}_{N'})}
\exp \left( -\frac{E_{IMN\xi}}{k_{\rm B} T}\right).\CR
\label{bare_su}
\enn
For the single rotational band with $(2I+1)^2$ fold degeneracy, 
levels of which the angular momentum are larger than $I_{\rm c}(T)
=\sqrt{6.2 \cJ k_{\rm B} T/\hbar^2}$ contributes only less than 10 percent 
of the partition function.
We call $I_{\rm c}(T)$ cut-off angular momentum.
When the band head of excited bands are much larger than the
energy of the cut-off level, that is,
\eq
u'(E^{\rm A}_N - E^{\rm A}_1) \gg \frac{\hbar^2 \{I_{\rm c}(T)\}^2}{2\cJ}, 
\quad (N\ge 7)
\label{cond_st}
\en
the higher bands do not have considerable occupation probabilities. 
Therefore, it is possible to describe the susceptibility 
with the ground band only.
Substituting energy of the ground band obtained through the 
Eqs. (\ref{ground_Coriolis}) and (\ref{E_eigen}) 
and the matrix element of $\hat{S_z}$ evaluated by 
the wave function of the ground band \Eq{WF_coriolis} into
\Eq{bare_su}, we get
\eqn
\chi &=& -\frac{1}{Z}\sum_{IMKi} \sum_{I'Ki'}
4\cJ \frac{\D_{ii'} S^2\frac{2I+1}{2I'+1}
|\bra IM10|I'M \ket |^2 |\bra IK10|I'K \ket |^2}
{I(I+1)-I'(I'+1)}\nonumber
\\ 
& & \hspace{6cm}
\times \exp \left( -\frac{\frac{\hbar^2(I(I+1)+S(S+1)+2KS)}{2\cJ}
+uE^{\rm A}_i}{k_{\rm B} T} 
\right).
\label{before_int}
\enn
This is different from the susceptibility for locked moment \cite{lockq}
only in the Coriolis term.
Since the average angular momentum is about $I \simeq 600$,
one can safely evaluate \Eq{before_int}, treating $I,K$ as a
continuous variable and replacing the sum by an integral.
Thus we get
\eqn
\chi &=& \frac{2\cJ}{3}\left( 
1-\frac{e^{-\beta S^2}}{S}\int_0^{S}e^{\beta t^2} dt 
\right)
=\sum_{n=0}^{\infty}\frac{2\cJ}{3\hbar^2}
\frac{(-1)^n 2^n}{(2n+3)!!}(\beta S^2)^{n+1} \CR
&=&\frac{2}{9}\frac{\hbar^2 S^2}{k_{\rm B} T}
\left( 1 -\frac{2}{5}\frac{\hbar^2 S^2}{2\cJ k_{\rm B} T} 
+\frac{4}{35}\left( \frac{\hbar^2 S^2}{2\cJ k_{\rm B} T} \right)^2 +\cdots
\right)
< \frac{2}{9}\frac{\hbar^2 S^2}{k_{\rm B} T},
\label{suscept_lock}
\enn
where $\beta = \frac{\hbar^2}{2\cJ k_{\rm B} T}$.
The leading order is the same as the locked moment susceptibility.
The higher orders mean the correlation by the Coriolis term;
in other words, a recoil effect due to 
the angular momentum of the super-electron spin, or Einstein-de Haas effect. 
We find from \Eq{suscept_lock} that
this effect reduces the magnetic susceptibility
compared with one of the locked moment.
This fact will be substantial in numerical calculation in
\Subsect{susceptIC}.
We close this subsection with evaluating the 
condition of neglecting the recoil effect.
For finite temperature, the cluster is excited by 
the thermal rotational excitation. 
For the rotational band with $(2I+1)^2$-fold degeneracy, 
the most major population of angular momentum is
$I_{\rm eff}(T)=\sqrt{\frac{2\cJ k_{\rm B} T}{\hbar^2}}$.
Therefore, the typical energy splitting by the Coriolis coupling becomes
$\hbar^2 \frac{S I_{\rm eff}(T)}{\cJ}$. We can neglect the recoil effect
under the high temperature condition
\eq
k_{\rm B} T \gg \frac{\hbar^2 S I_{\rm eff}(T)}{\cJ}.
\label{neg_Coriolis}
\en
\subsection{Susceptibility for Weak Coupling}
\label{susceptWC}

In this subsection we discuss the susceptibility for
the weak coupling and high temperature limits.
Firstly we discuss the energy eigenvalue and the eigenfunction of 
$H_{\rm rot}+H_{\rm coupl}$ in the weak coupling.
The angular momentum of rotor $R$ is approximately conserved due to 
the weak coupling.
In the absence of coupling, the energy eigenvalue is a single rotational
band with $(2R+1)^2$ fold degeneracy.
The coupling removes the degeneracy; the first order energy shift 
is determined by solving the secular equation in degenerate space.
Therefore, the energy eigenvalue is described as 
$\frac{\hbar^2 R(R+1)}{2\cJ}+\Delta E_{I\lm(R)}$
where $\lm(R)$ specifies states adiabatically connected with $R$, 
and $\Delta E_{I\lm(R)}$ stands for 
the eigenvalue of the weak anisotropic coupling.
The unperturbed base is expressed by
\eq
|IMRk \ket = \sum_{\mu \sigma} \bra R \mu S \sigma |IM \ket
\cD_{\mu k}^R |S \sigma \ket
\en
The coupling is treated as the first order perturbation to degenerate
states. The perturbed wavefunction is expressed as
\eq
\label{weak_base}
\Psi_{\lm(R) I M}=\sum_{k\sigma}
 \left\langle R \mu S \sigma | I M \right\rangle
 {\cal D}^R_{\mu k}(\Omega)c^{\lm(R)}_{k}\left| S \sigma
\right\rangle, 
\en
where
the coefficient $c^{\lm(R)}_{k}$ is determined by the diagonalization
of $H_{\rm coupl}$ in each degenerate rotational level.

Now we evaluate the magnetic susceptibility for weak coupling
by employing the \Eq{par2}.
\Equation{par2} reduces the calculation of magnetic susceptibility
to the evaluation of second order energy shift.
The energy denominator between different $R$ levels are much larger than
the one of the same $R$ in the weak coupling.
So, we neglect the contribution from different $R$ in the perturbation.
The orthogonality condition
$\sum_{k} c^{\lm(R)\ast}_{k}c^{\lm'(R)}_{k}
=\delta_{\lm \lm'}$ yields the matrix element of
Zeeman interaction between the weak coupling base, \Eq{weak_base}, as
\eqn
\label{wm_matrix}
 \left\langle\Psi_{\lm(R) IM}|\hat{S_z}|\Psi_{\lm'(R) I'M}\right\rangle
  &=& \delta_{\lm,\lm'}\sqrt{2I'+1}
 (-1)^{I'}\left\langle I'M10|IM \right\rangle \CR
&&\times 
\sqrt{S(S+1)(2S+1)}(-1)^{R-S+1}W(ISI'S;R1).
\enn
Under the approximation, \Eq{par2} becomes
\begin{equation}
\chi
=-\frac{2}{Z}\sum_{R}\left( \sum_{\lm(R) M I \neq  I' }
\frac{|\left\langle\Psi_{\lm(R) IM}|\hat{S_z}|\Psi_{\lm(R) I'M}\right\rangle |^2}
{\Delta E_{I \lm(R)}-\Delta E_{I' \lm(R)}}
e^{-\frac{E_{R}+\Delta E_{I' \lm(R)}}{k_{\rm B} T}}\right),
\label{par4}
\quad E_R = -\frac{\hbar^2 R(R+1)}{2\cJ}.
\end{equation}
Since $\Delta E_{I' \lm(R)}$ is small because of the weak coupling,
we can expand $e^{-\frac{\Delta E_{I' \lm(R)}}{k_{\rm B} T}}$
up to first order as $1-\frac{\Delta E_{I' \lm(R)}}{k_{\rm B} T}$.
Rearranging the summention, we get
\eq
\chi=-\frac{2}{Z}\sum_{R}\left( \sum_{\lm(R) M I>I' }
\frac{|\left\langle\Psi_{\lm(R) IM}|\hat{S_{\rm z}}
|\Psi_{\lm'(R)I'M}\right\rangle |^2}
{k_{\rm B} T}
\right)e^{-\frac{E_R}{k_{\rm B} T}}.
\label{Tover1form}
\en
Putting \Eq{wm_matrix} into \Eq{Tover1form}, and expanding for $R$,
Therefore, we obtain the magnetic susceptibility
\eqn
\chi&=&
2\frac{\sum_{R=0}^{\infty}\left( \frac{4S(S+1)(2S+1)R^2}{9}+O(R) \right)
\exp\left(-\frac{\hbar^2 R(R+1)}{2\cJ k_{\rm B} T}\right)}
{\sum_{R=0}^{\infty} \left( (2R+1)^2(2S+1) \right)
\exp \left( -\frac{\hbar^2 R(R+1)}{2\cJ k_{\rm B} T}\right)}
\label{sus_W6}
\enn
At high temperature limit many levels contribute to the summention over $R$.
Therefore one can evaluate \Eq{sus_W6}, treating $R$ as a continuous
variable and replacing the sum by integral. Thus, we get
\eq
\chi = \frac{2S(S+1)}{9k_{\rm B} T} 
\en
It is observed in this subsection that the locked moment susceptibility is
obtained in the weak coupling and high temperature limit.
One should note that this does not mean the super-electron spin is locked 
in the weak coupling.
In fact, taking into account the axial deformation,
the susceptibility of weak and strong coupling becomes different.
In the weak coupling, the energy level is dependent on the axial deformation
through the $k$ quantum number:
\eq
E_{IMRk}=\frac{\hbar^2}{2\cJ_1}(R(R+1)-k^2)+\frac{\hbar^2}{2\cJ_3}k^2.
\en
The matrix element is independent of the deformation, because the
\Eq{wm_matrix} is independent of $\lm$.
Therefore, the leading order of $1/k_{\rm B} T$ 
expansion of magnetic susceptibility 
is deformation independent even if the energy is dependent on $k$,
and is $2S(S+1)/9$ even in an axial deformed cluster.
On the other hand, in strong coupling the energy level for the ground band
expressed as
\eq
E_{IMK}=\frac{\hbar^2}{2\cJ_1}(I(I+1)-K^2+KS)+\frac{\hbar^2}{2\cJ_1}K^2
\en
If we neglect the Coriolis term, this is the same as the
one of the weak coupling, apart from the degeneracy.
The matrix element for axial deformed cluster is not different 
even in the strong coupling.
Hence, the magnetic susceptibility of the axial deformed cluster
is same as \Eq{before_int} apart from the energy. 
Since the matrix element divided by the energy difference 
$I(I+1)-I'(I'+1)$ is dependent on the deformation,
the magnetic susceptibility becomes deformation dependent.
Therefore, considering the axial deformation,
the susceptibility for the weak and strong coupling limits are different.
The two limits happen to be the same value when the value of $\cJ_i's$ 
are equal.
This equality does not mean the super-electron spin is locked 
in the weak coupling.

\subsection{Intermediate Coupling}
\label{susceptIC}
We numerically calculate the magnetic susceptibility for intermediate coupling
in this subsection.
In numerical calculations of the deflection profile and the magnetization,
a main task is diagonarization of Hamiltonian matrices of large dimensions.
In the source area, the partial Hamiltonian $H_{\rm mag}$ is zero, and
therefore the total angular momentum is conserved.
The dimension of matrices to be diagonarized is approximately 
$S \times I_{\rm eff}(T)$.
The dimension of matrices to be diagonarized is of
the order of 10$^{5}$.
In the magnetic field where states having $I/2$ different are mixed,
the number of dimension is magnified by $I/2$,
and becomes eventually of the order of 10$^{7}$.
It may not be feasible in numerical calculations.
Let us use a similarity transformation in order to scale down
the angular momenta.
In \Eq{c_matrix}, the matrix element is approximated as
\begin{equation}
\label{c_appro}
  (H_{\rm coupl})_{RK,R^{\prime}K^{\prime}} \simeq 
  A^{\prime}_\kappa \frac{\sqrt{(2R+1)(2R^{\prime}+1)}}{2\overline{R}+1}
  d_{ \kappa \mu}^{4}(\theta_{1})
  d_{  \mu 0}^{4}(\theta_{2})
\end{equation}
with
\begin{equation}
  \overline{R} = \frac{R+R^{\prime}}{2}~~~{\rm and}~~~\mu = R^{\prime} - R,
\end{equation}
and two angles,$\theta_{1}$ and $\theta_{2}$, are
defined respectively as
\begin{eqnarray}
\cos \theta_{1} & = & \sqrt{\frac{\overline{K}}{\overline{R}}}
~~~~ {\rm with} ~~~~\overline{K}=\frac{K + K^{\prime}}{2},
\nonumber
\end{eqnarray}
and
\begin{eqnarray}
\cos \frac{\theta_{2}}{2} & = & 
 \sqrt{\frac{(\overline{R}-S+I)(S - \overline{R} +I)}{4\overline{R}S}}.
\end{eqnarray}
These angles are invariant under a scale transformation defined
as
\eq
 \label{scale_t}
\pmatrix{
    r,r^{\prime} ~\overline{r} \cr
 k,k^{\prime} ~\overline{k} \cr
 s,i,m \cr
}
= \eta
\pmatrix{
    R,R^{\prime} ~\overline{R} \cr
 K,K^{\prime} ~\overline{K} \cr
 S,I,M \cr
}.
\en
In a similar manner, the matrix elements given in \Eq{m_matrix}
are expressed approximately as
\begin{eqnarray}
\label{m_simul}
 \left\langle\Psi_{\nu \pi_K IM} \mid H_{\rm mag}
 \mid\Psi_{\nu'I'\pi_K M}\right\rangle
 &  \simeq &
  Bg_{\rm s} \sqrt{S(S+1)} \frac{\sqrt{(2I+1)(2I^{\prime}+1)}}{2 \overline{I}}
 \nonumber
 \\
  & & \sum_{RK} f_{RK}^{\nu I *} f_{RK}^{\nu^{\prime} I^{\prime}}
  (-)^{I'-I}
  d_{0 \, I-I^{\prime}}^{1}(\theta_{3})d_{I-I^{\prime}\,0}^{1}(\theta_{4})
\end{eqnarray}
with
\begin{eqnarray}
\cos \theta_{3} & = & \sqrt{\frac{M}{\overline{I}}}
\nonumber
\\
\cos \frac{\theta_{4}}{2} & = & 
 \sqrt{\frac{(\overline{I}-S+R)(S - \overline{I} +R)}{4\overline{I}S}}
\end{eqnarray}
The two additional angles, $\theta_{3}$ and $\theta_{4}$ are also
invariant under the scale transformation.
Eventually these four angles are invariant with respect to
the scale transformation in \Eq{scale_t}.
The four $d$-functions are all smooth functions of the four angles.

In the numerical calculation we scale down the super-electron spin 
quantum number $S$.
To make $\hat{S}^2$ invariant under the similarity transformation, 
\Eq{scale_t}, the Plank constant is changed according to $\eta$.
In other words, we can scale down the angular momentum quantum
number by adjusting Plank constant as 
\eq
\hbar'=\frac{\hbar}{\eta},~~~ s=\eta S.
\label{reduction}
\en
The temperature $k_{\rm B} T$, coupling $u'$, and magnetic field parameter
$g_{\rm s} B \hbar S$, which have dimension of energy, are invariant 
under the similarity transformation.
In the following, the unit of these energy parameter is selected as 
$\frac{\hbar^2 S^2}{2\cJ}$ which is also invariant under the 
similarity transformation.

In the numerical calculation, we set the cut off of super-spin 
$s=S_{\rm c}$ and the cut off of total angular momentum $I_{\rm max}$.
$\eta$ is decided through the $S_{\rm c}$ and \Eq{reduction} as
$\eta=S_{\rm c}/S$.

\begin{figure}[htdp]
\centerline{\psfig{figure=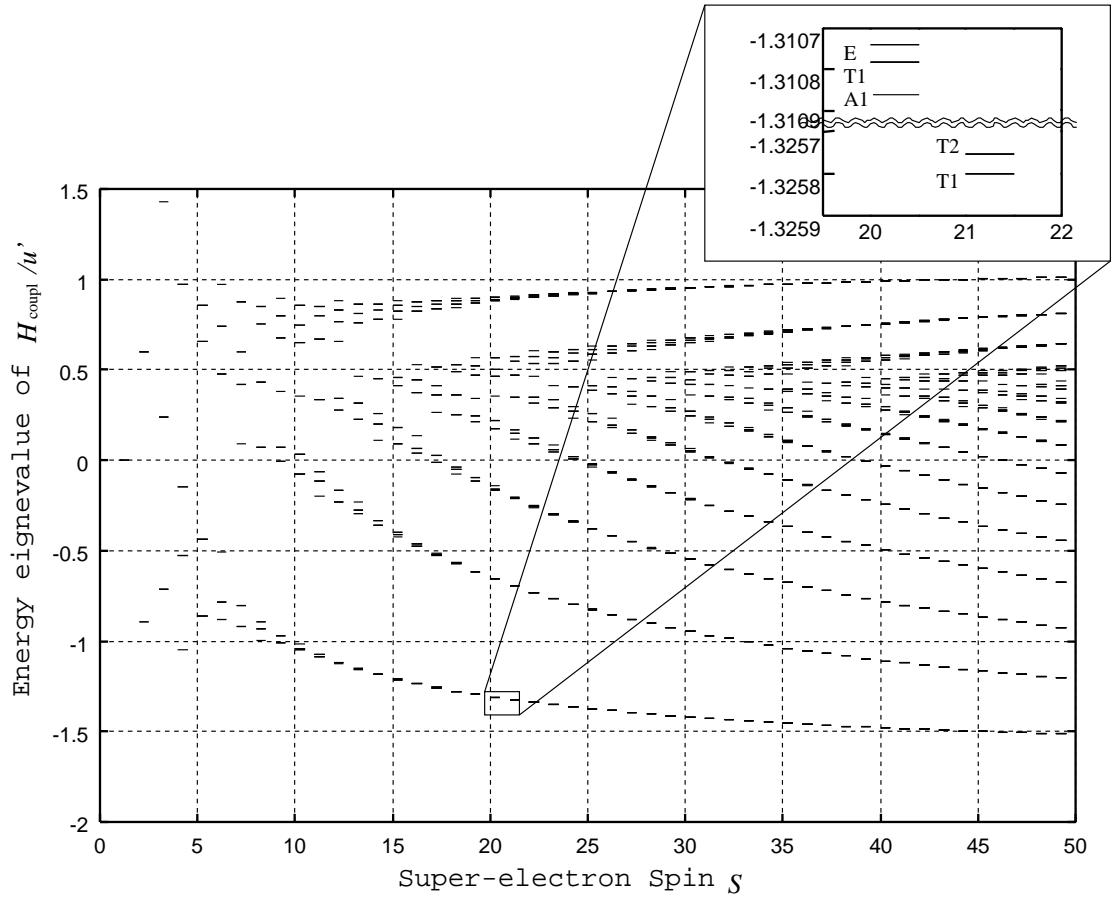,height=12cm}}
\caption{The eigenvalue of $H_{\rm coupl}/u'$ as a function of the magnitude of Spin $S$.}
\label{u_energy}
\end{figure}

We are now ready to calculate numerically 
the susceptibility for intermediate coupling.
It is already known by the analysis in the above two subsections that 
the susceptibility becomes the locked moment in the limits.
To achieve locked moment limit, the ground state of the
$H_{\rm coupl}$ has to be approximately six fold degenerate.
When we diagonalize $H_{\rm coupl}$ in $S_{\rm c}=10$
(see \Fig{u_energy})
three energy levels having 1-fold, 3-fold, and 2-fold degeneracy
from lower energy to higher make a bunch around the ground state.
The energy differences of these states are about 0.01.
In contrast to this, we find from \Fig{u_energy} that 
the energy difference between bunch of the ground state and 
the first excited bunch is about 1.0.
The energy spacing between six states around the ground states are
much smaller than the energy spacing between the ground bunch and 
the first excited bunch.
Therefore, it seems reasonable that 
the ground state is approximately 6-fold degenerate for $S_{\rm c}=10$.

A truncation of the angular momentum is necessitated due to 
computational limitations.
We select $I_{\rm max}=80$ in the calculation. 
As mentioned previously, we do not confident
about the calculation of partition function out of the reliable range
of temperature.
Hence, the numerical calculation of partition function is reliable only 
$I_{\rm c}(T)\leq I_{\rm max}$. This restricts the range of temperature 
as $k_{\rm B} T < 20.6$. We choose $20$ for the temperature.  

\begin{figure}[htdp]
\centerline{\psfig{figure=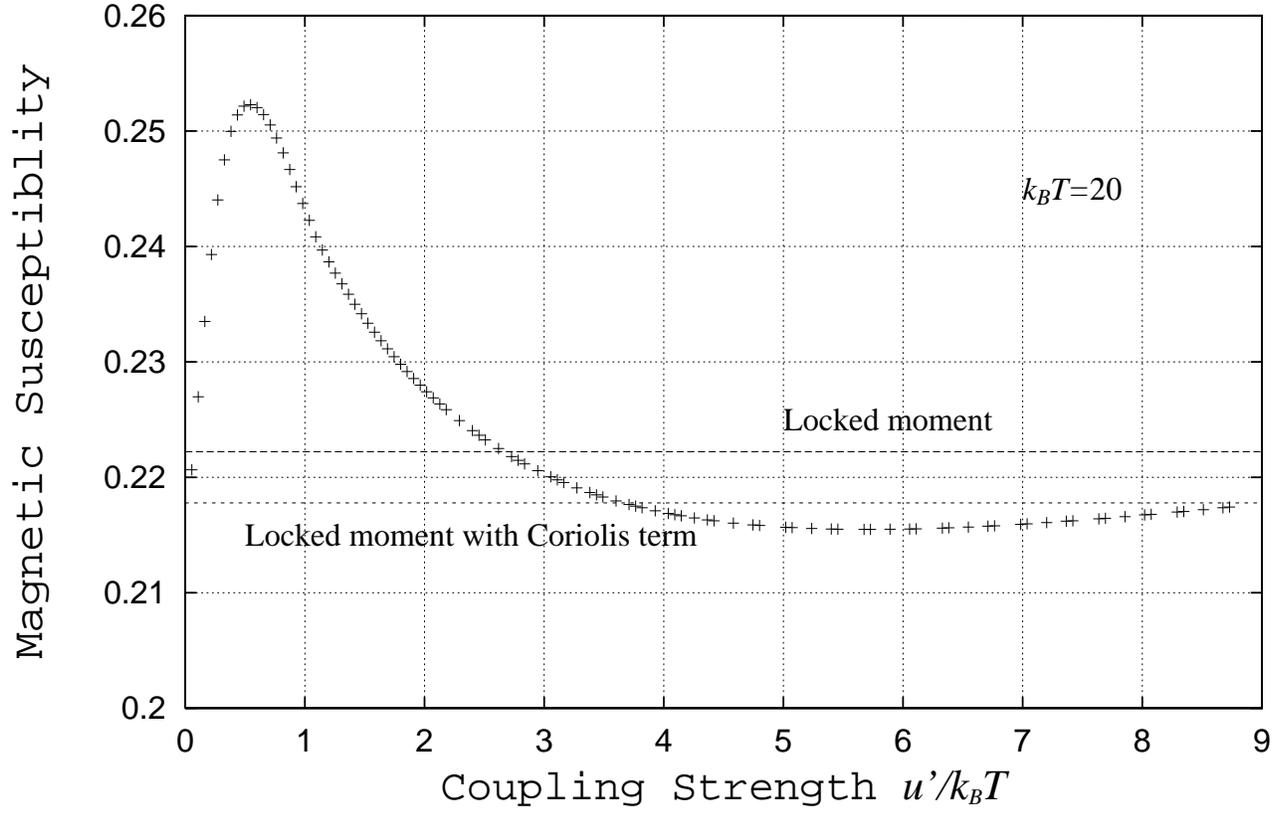,height=12.0cm,angle=-90}}
\caption{The magnetic susceptibility calculated by the perturbation theory
of the intermediate coupling model as a function of the ratio of coupling
strength to temperature.}
\label{suscept}
\end{figure}

We display the magnetic susceptibility of our calculation 
in \Fig{suscept}.
The first point to be discussed is whether our calculation reaches
the two distinct limits discussed above.
If the coupling is much smaller than the temperature, the susceptibility
decreases and converges to $2/9$.
Therefore our calculation agrees with the
analysis for weak coupling discussed in \Subsect{susceptWC}.

Let us discuss the strong coupling limit.
In \Subsect{susceptSC} we discuss the condition of temperature
and coupling in which the susceptibility attains the locked moment value.
The condition for the coupling and for the temperature 
are expressed in \Eq{cond_st} and \Eq{neg_Coriolis}, respectively.
These two conditions for $S_{\rm c}=10$ become;
\eq
\frac{u'}{k_{\rm B} T} \gg 3.1,~~{\rm and}~~
k_{\rm B} T \gg 4.
\label{cond_10}
\en
Therefore, we can regard $\frac{u}{k_{\rm B} T} \simeq 9$ 
for $k_{\rm B} T=20$ as the strong coupling region.
However, in \Fig{suscept}, the susceptibility is not $2/9$ even if
the coupling becomes larger than 9.
In particular, one might not readily believe that the susceptibility 
is smaller than $2/9$.
One should recall that we evaluate the strong coupling limit of 
susceptibility including the effect of Coriolis coupling 
as \Eq{suscept_lock}.
We find from this expression that the susceptibility is 
decreased by the recoil effect.
We calculate the susceptibility for $k_{\rm B} T=20$ 
through \Eq{suscept_lock}.
The result is illustrated in \Fig{suscept} and is close to the
numerically computed susceptibility. 
Nevertheless, the numerically computed susceptibility shows some deviation
from the value of \Eq{suscept_lock}.
A possible reason for this is that the quantum fluctuation 
between different potential minima of the coupling
decreases the susceptibility.

We find a peak at $\frac{u'}{k_B T}=0.55$ in \Fig{suscept}.
This peak is accounted that the energy splitting of 
anisotropic interaction $2u'$ is comparable to
the energy splitting by Coriolis term
$\frac{\hbar^2}{2\cJ}4I_{\rm eff} S$.
The width of energy splitting of anisotropic interaction corresponds
to the energy difference between potential maxima and minima i.e.
$\frac{8}{3}u'$. In practice the width becomes smaller due to the
quantum fluctuation and 
is found $2u'$ for $S=10$ from \Fig{u_energy}.
Then the peak is expected to be,
\eq
2u' \simeq \frac{\hbar^2}{2\cJ}4I_{\rm eff} S ~~ \Leftrightarrow ~~
\frac{u'}{k_{\rm B} T} \simeq 2\sqrt{\frac{\hbar^2 S^2}{2\cJ}
\frac{1}{k_{\rm B} T}} \simeq 0.45 .
\label{peak_cond}
\en
In fact, this condition is consistent with the observed position of peak.

Let us move to the next point. As mentioned in introduction, 
one often assumes the superparamagnetism, in which susceptibility is 
$\frac{1}{3}S(S+1)$, for the clusters to analyze the experiments.
From \Fig{suscept} that the superparamagnetic limit is not reach in any
range of coupling strength.

The last point is the temperature dependence of the magnetization.
It is reasonable to consider that 
the susceptibility decreases as the temperature increased
because of thermal fluctuations.
A quite opposite dependence is reported in Ref. \cite{fe}.
However, our calculations does not show such a behavior in any range of
coupling.
\section{Profiles and Magnetization}
\label{Pro_Mag}
%
%

As mentioned in the previous section, a truncation of the angular momentum
is necessary due to the computational limitations.
The procedure of calculating profile includes the large dimensional
diagonalzation of Hamiltonian $H_{\rm mag}$. To make it practicable,
we choose a truncated value of $I_{\rm max}=26$, and the magnitude
of super-electron spin $S_{\rm c}=10$.
As discussed in section \ref{Mag_Sus}, 
this choice narrows the reliable range of temperature.
The reliable range of temperature for this cut-off becomes $k_{\rm B} T <2$.
Since high temperature is one of the conditions for the strong coupling
limit, we should select as large temperature as possible.
Eventually, we choose $k_{\rm B} T =2$ which is 
maximum for the reliable range of $I_{\rm max}=26$.

\begin{figure}[hdtp]
\psfig{figure=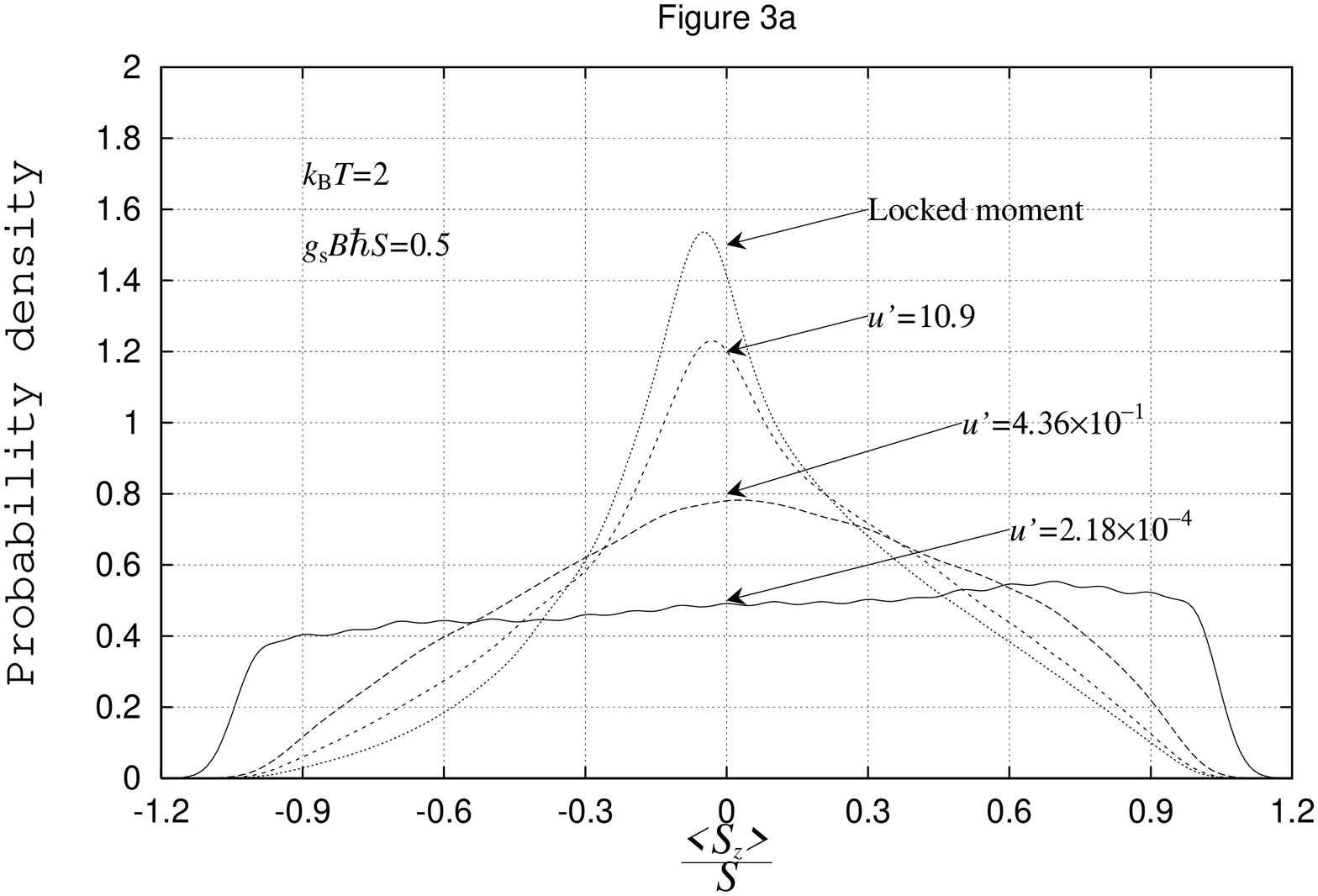,height=10.0cm,angle=-90}
\psfig{figure=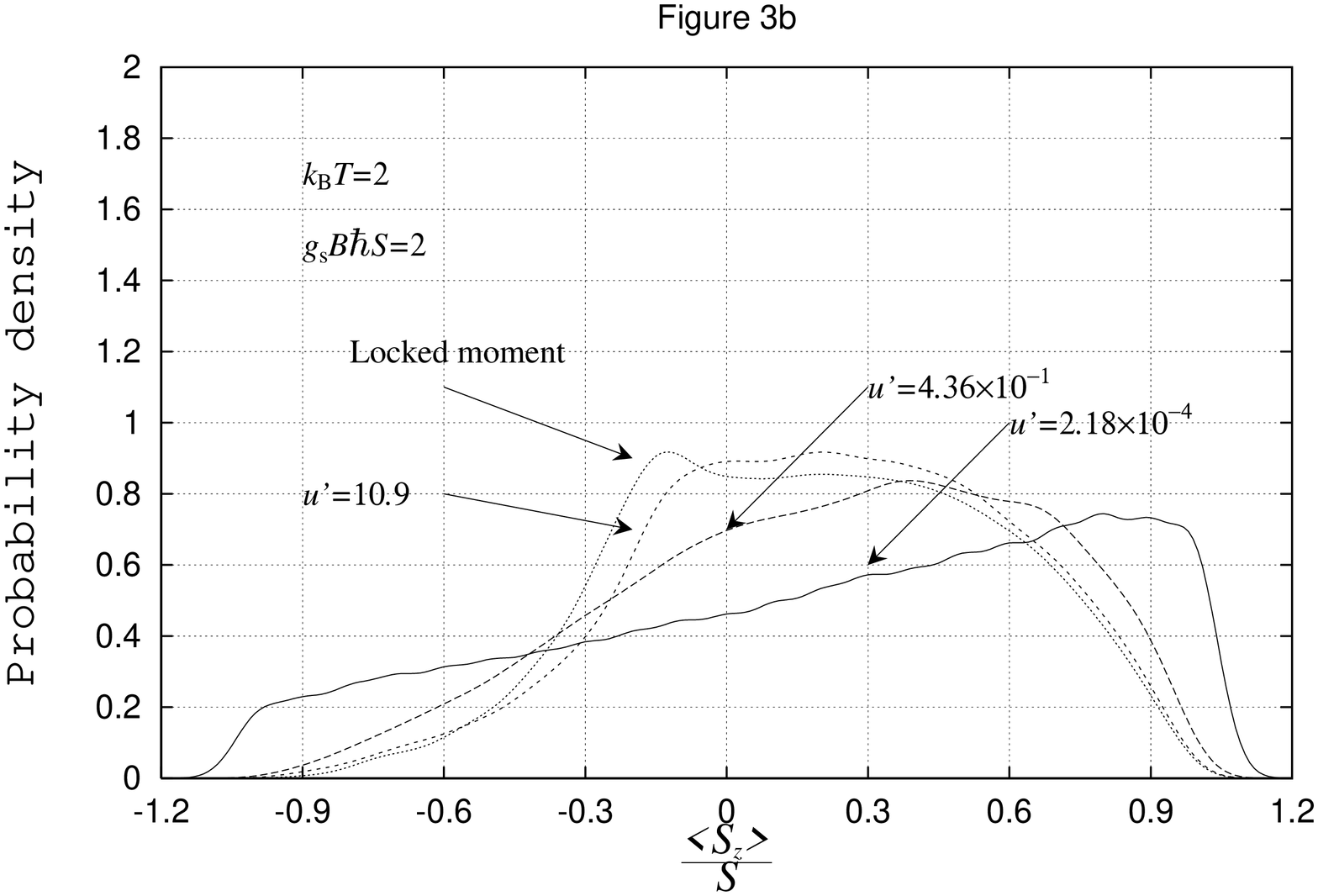,height=10.0cm,angle=-90}
\psfig{figure=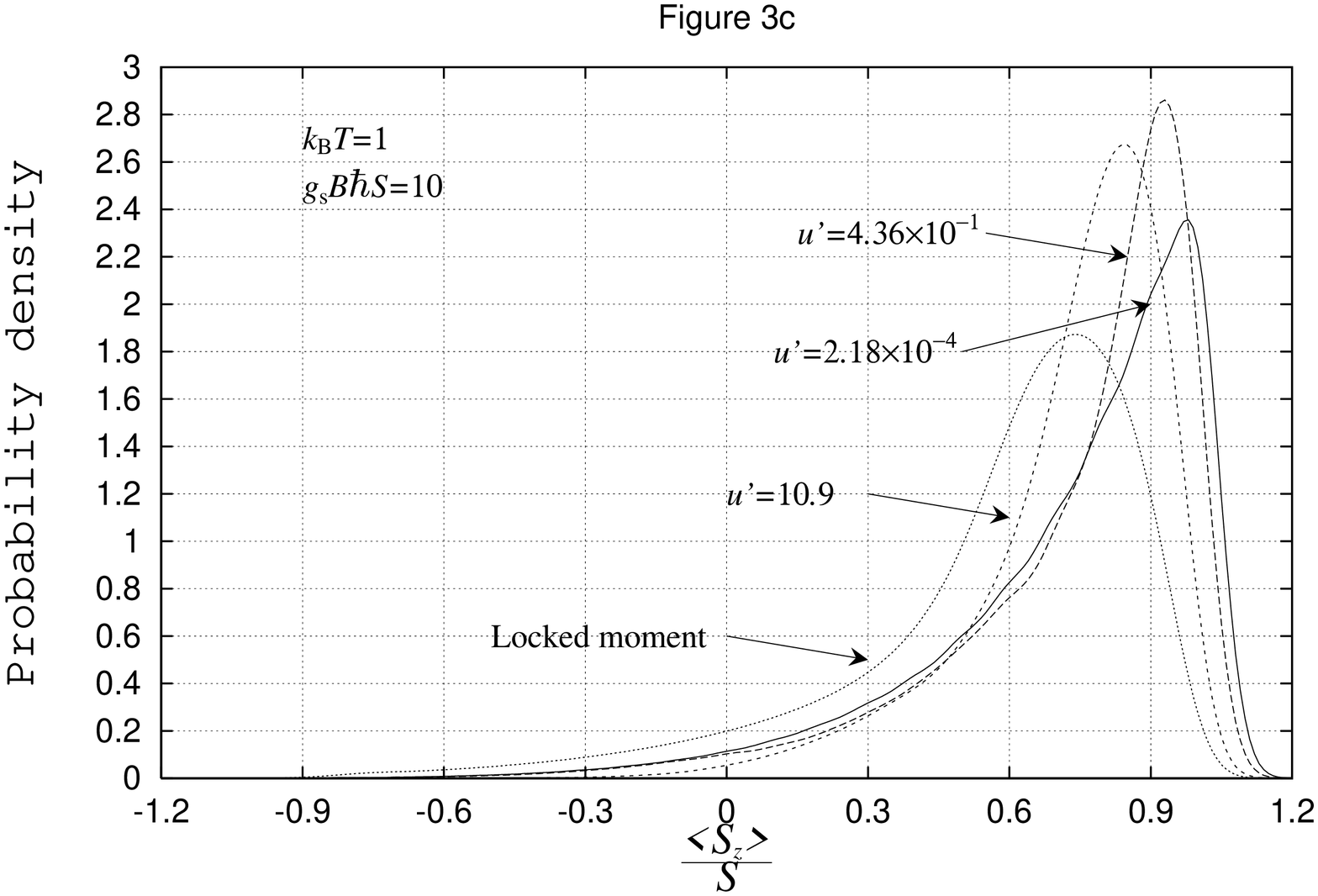,height=10.0cm,angle=-90}
\caption{The profile for three coupling strength
$u'=2.24 \times 10^{-4}$,$u'=4.47 \times 10^{-1}$,
and $u'=11.2$.
of magnetic field $g_{\rm s} B\hbar S=0.5,2,10$.
The temperature is set to $k_{\rm B} T=2$ for $g_{\rm s} B\hbar S=0.5,2$ 
and to $k_{\rm B} T=1$ for $g_{\rm s} B\hbar S =10$.}
\label{mag_dep}
\end{figure}

Firstly we pay attention to the magnetic field dependence of the deflection
profile. Figure \ref{mag_dep} shows the deflection profile
for three different strengths of the magnetic field 
$g_{\rm s} B \hbar S =0.5,2,10$,
respectively.
Many experiments show that peak deflection is
always to the strong field.
Our calculation also exhibits deflection of a similar
nature. By solving the classical equation of motion,
De Heer et al. have demonstrated in Ref \cite{relaxC} that the coupling
between super-spin and cluster body causes 
the deflection behavior.
We will have a closer examination of this statement 
in terms of our quantal model.

\begin{figure}[htdp]
\centerline{\psfig{figure=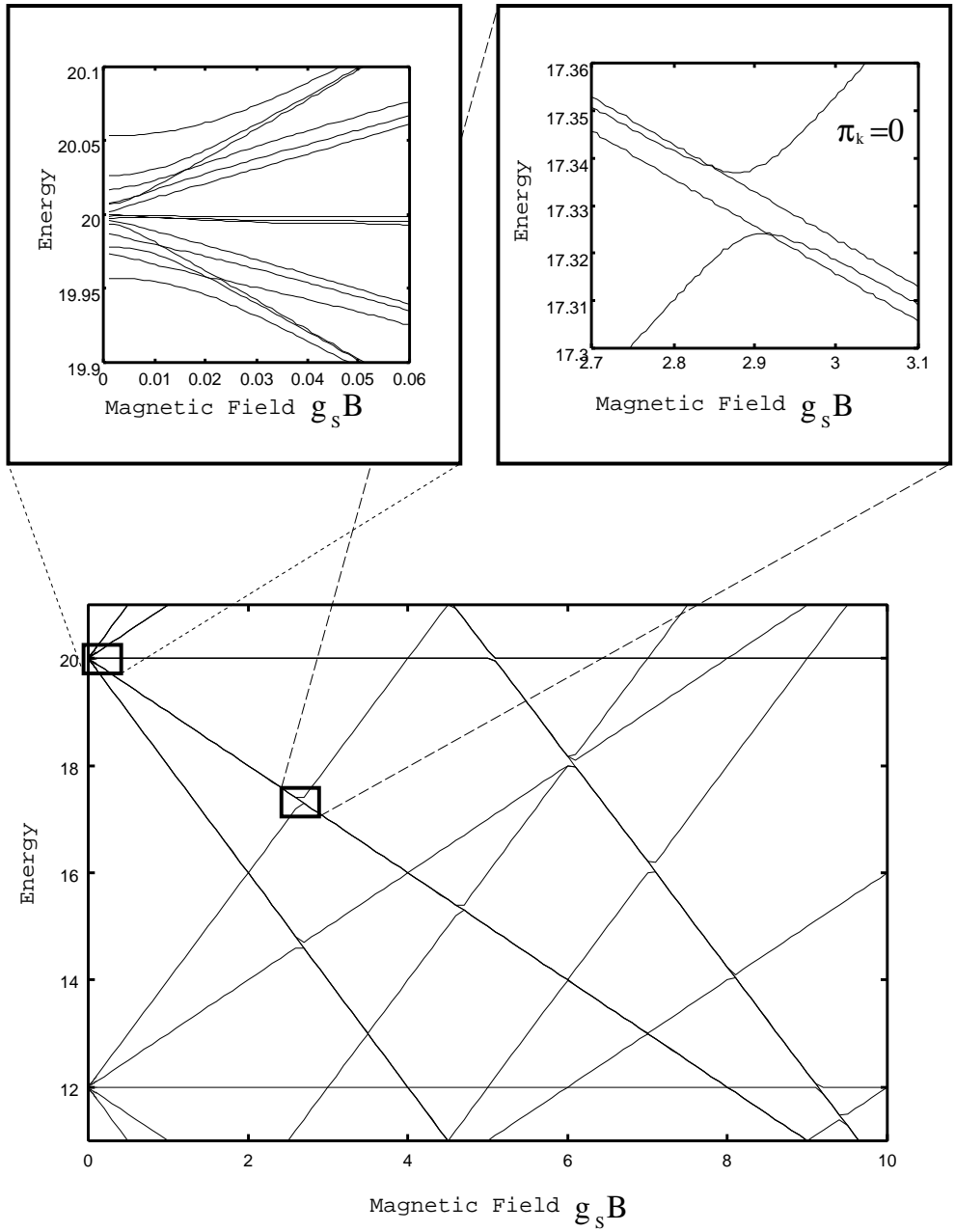,height=17.0cm}}
\caption{Energy levels of the weak coupling ($u'=4.47 \times 10^{-3}$)
as a function of magnetic field for $S_{\rm c}=2, M=0,$ 
and $R \simeq 3$ or $4$.}
\label{enlv}
\end{figure}

In order to make our discussion transparent, we discuss 
the weak coupling limit first.
We can see from \Fig{mag_dep} that the profile of the weak coupling
is a flat but slightly sloping distribution in the weak magnetic field.
\Figure{enlv}, which shows the energy levels of 
the weak coupling $u'=4.47 \times 10^{-3}$ for 
$S_{\rm c}=2, M=0, \pi_k=0$ and $R \simeq 3$ or $4$, helps us
to understand the distribution in terms of the energy level.
In the absence of magnetic field, the weak coupling
removes 15-fold degeneracy of unperturbed rotational levels; 
Energy levels result in a bunch of states around 
the unperturbed rotational levels.
The applied magnetic field further splits these levels, and
rearranges each one more likely to be the eigenfunction of $S_z$,
when the applied magnetic field becomes stronger than the coupling.
In other words, the super-spin precesses about the direction of magnetic
field independently from the cluster (decoupling)
for $g_{\rm s} B \hbar S> u'$.
Although the magnetization increases proportional to the magnetic field before
the decoupling, occupation probabilities of the same $R$ levels become 
almost equal because of the weak coupling.
Then, the magnetization remains small and the profile becomes
flat, rectangle like distribution.
When the Zeeman splitting equals to the typical energy difference of the
rotational levels: $2g_{\rm s} B \hbar S = 2\hbar^2 R_{\rm eff}/2\cJ$,
the pseudo-crossing between different rotational levels occurs.
To put it another way, the pseudo-crossing occurs where the Larmor precession
frequency $\omega_{\rm L}$ is comparable 
to the cluster rotation frequency $\omega_{\rm rot}$; 
$\omega_{\rm L} \simeq \omega_{\rm rot}$.
That is to say, 
the pseudo-crossing leads to exchange the occupation probabilities 
between those levels.
Therefore, the magnetization increases and profile develops a peak at
$\bra S_z\ket /S=1$ like the Boltzmann distribution.

\begin{figure}[htdp]
\centerline{\psfig{figure=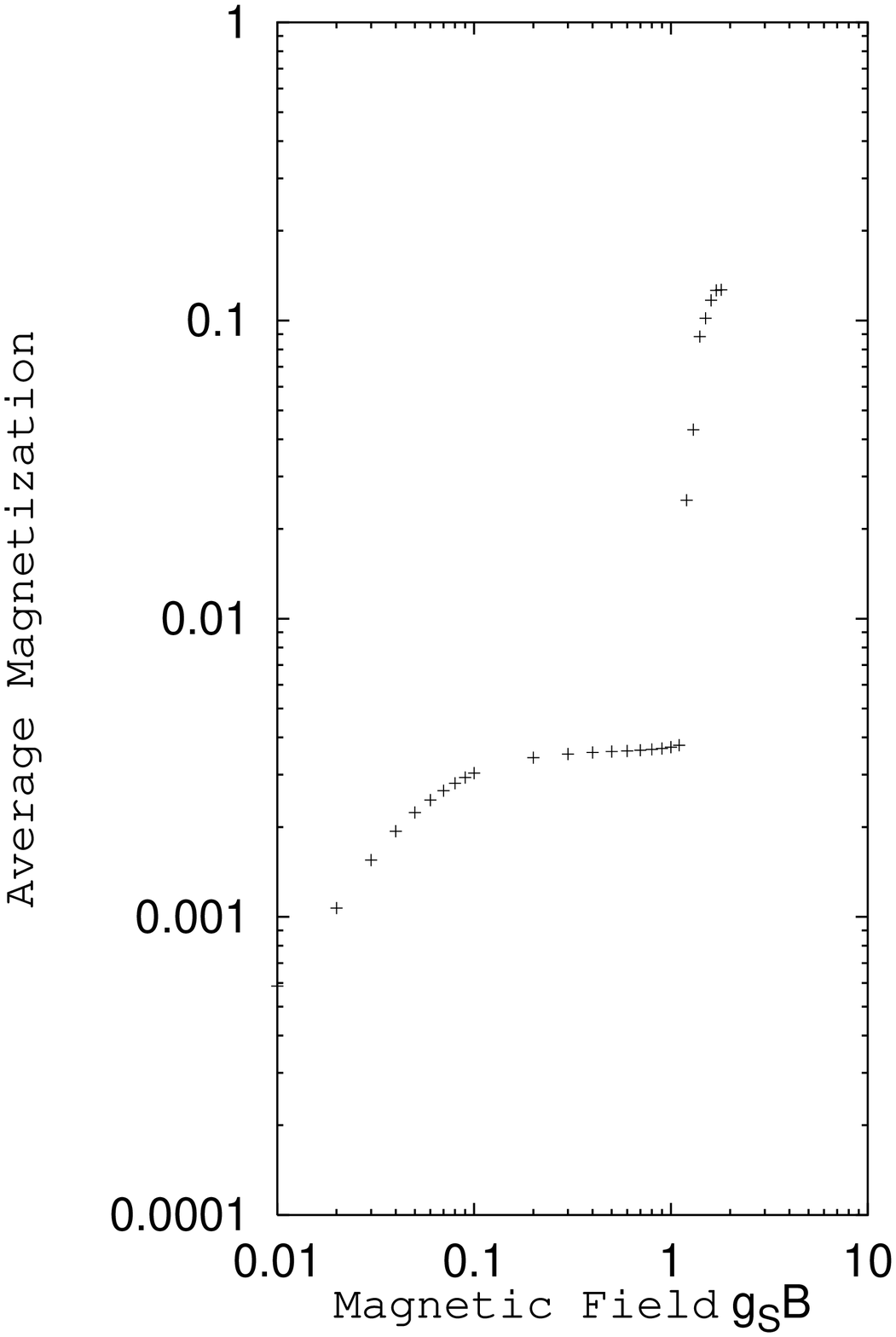,height=15.0cm}}
\caption{An example of peculiar behavior of magnetization for $S_{\rm c}=2$}
\label{anormalous_dep}
\end{figure}

The above discussion leads us to divide the mechanism of magnetization
into two types, that is, the magnetization by the process of decoupling
and pseudo-crossing.
When the coupling is weak, these two are separated definitely.
This makes the magnetic field dependence of magnetization peculiar as 
shown in figure \ref{anormalous_dep}.
The magnetization linearly increases as the magnetic field 
in the process to decoupling;
then, remains steady by the decoupling of super-spin; 
finally, increases suddenly because of the pseudo-crossing.
Since the magnetic field region of the decoupling and pseudo-crossing can not
be divided into the intermediate and strong coupling, the peculiar behavior
of magnetization disappears.

Let us now turn to the strong coupling.
The calculated susceptibility (Fig \ref{suscept}) indicates that
$u'=11.2$ is in the strong coupling region.
However, the calculated profile for $u'=11.2$ 
in figure \ref{mag_dep} is not identical with the locked moment profile.
We attribute this to the Coriolis term, which is important at the
temperature of our ensemble, $k_{\rm B} T=2$. A higher temperature was possible
in the calculation in ref \cite{VB96} because simpler anisotropy term
there permitted smaller dimension Hamiltonian matrices.

Finally, let us move to the discussion of the profiles 
in intermediate coupling.
We characterize the profiles by the cumulants.
Since the cumulant higher than the second order vanishes 
for the Gaussian probability distribution, 
the characterization of the profiles by cumulant clearly shows us
the deviation from Gaussian profile.
In our calculation the third and the fourth order cumulants 
are much smaller than the first 
and the second one,
so the Gaussian profile gives a reasonable description.

\begin{figure}[htdp]
\centerline{\psfig{figure=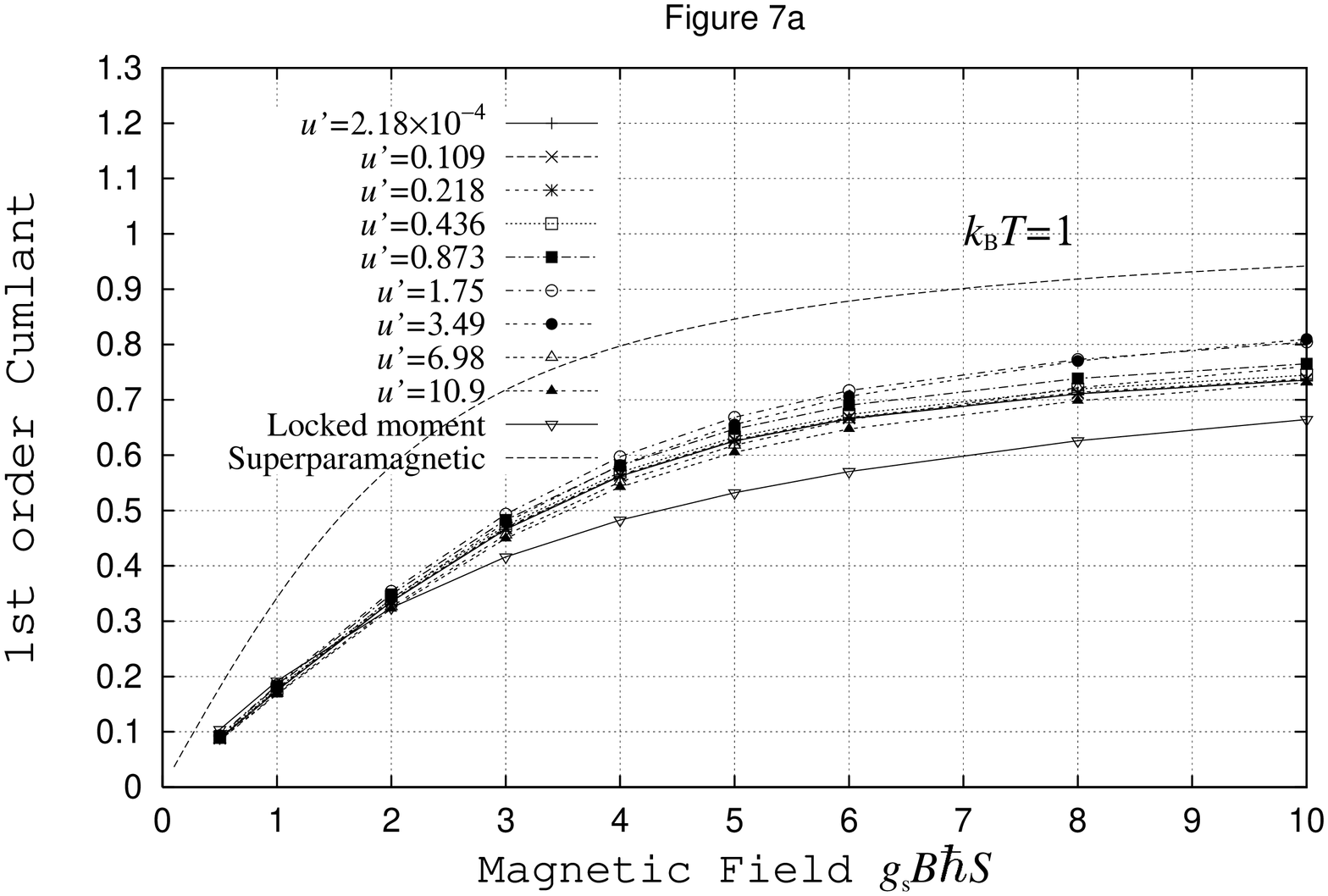,height=10.0cm,angle=-90}}
\centerline{\psfig{figure=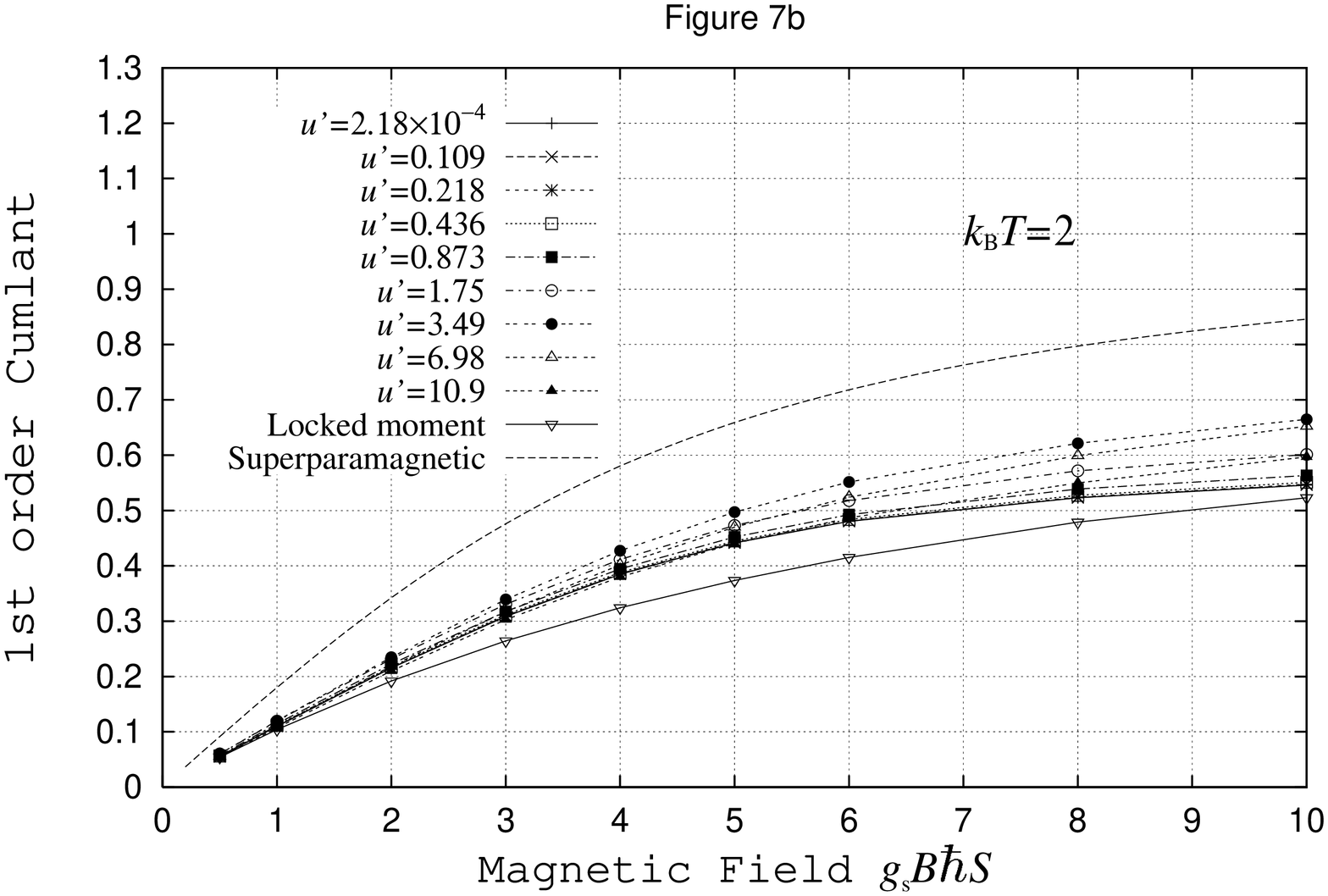,height=10.0cm,angle=-90}}
\caption{The first order cumulant of each profile 
as a function of the ratio of magnetic field. The temperature is fixed at
$k_{\rm B} T=1$ and for $k_{\rm B}T=2$ for Fig. 7a and Fig. 7b respectively.}
\label{cumt}
\end{figure}

\begin{figure}[htdp]
\centerline{\psfig{figure=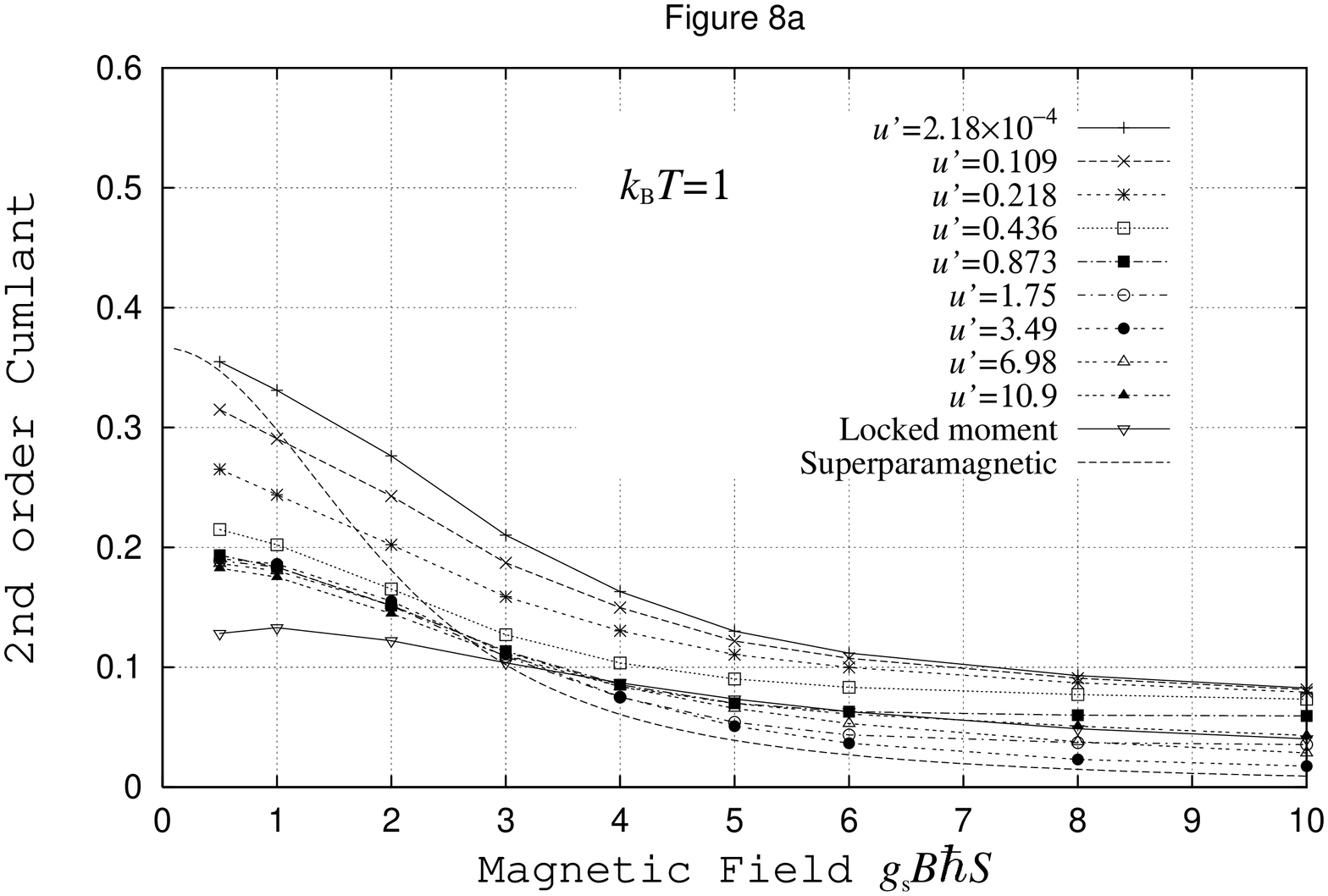,height=10.0cm,angle=-90}}
\centerline{\psfig{figure=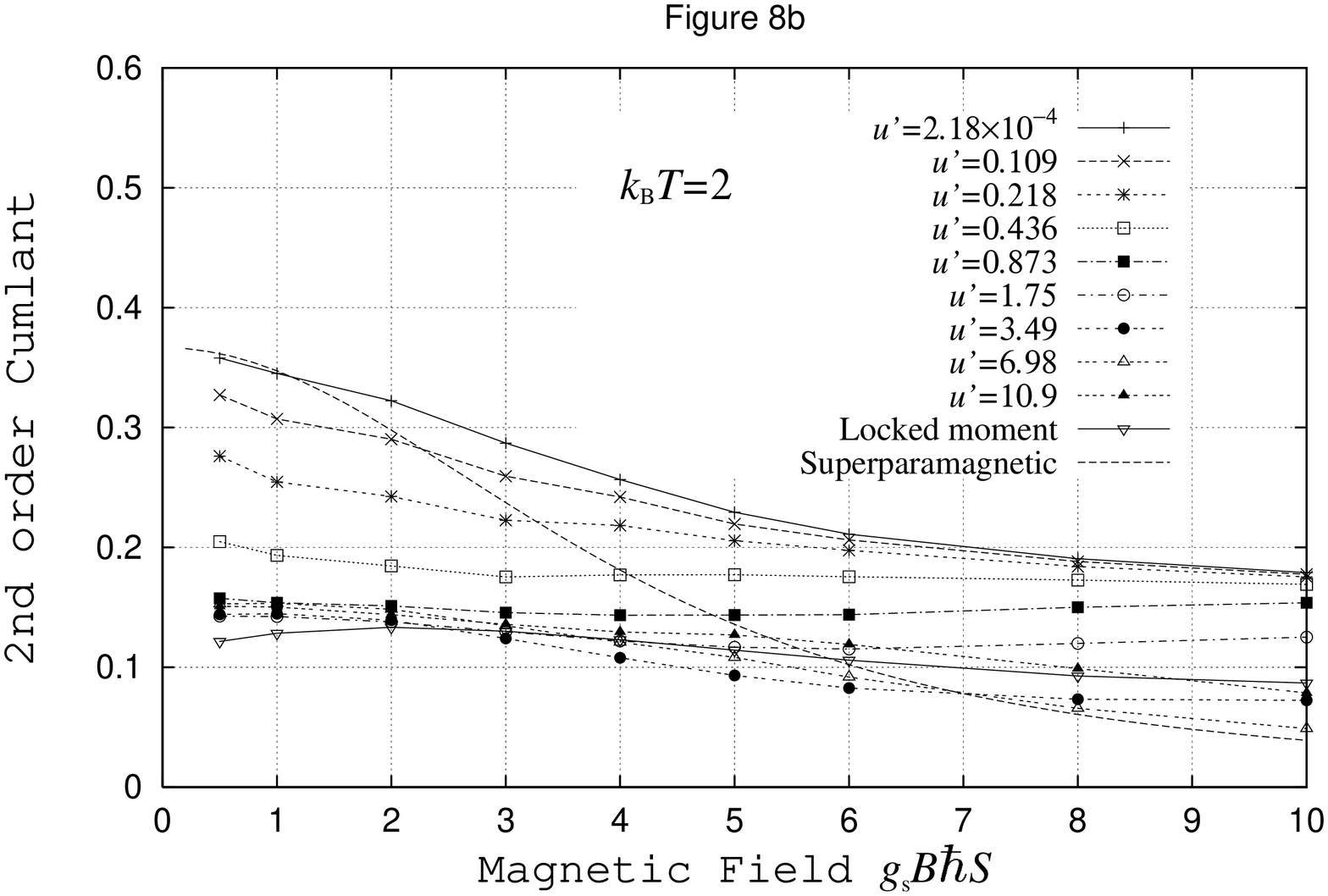,height=10.0cm,angle=-90}}
\caption{The second order cumulant of each profile 
as a function of the ratio of magnetic field. The temperature is fixed at
$k_{\rm B} T=1$ and for $k_{\rm B}T=2$ for Fig. 8a and Fig. 8b respectively.}
\label{2cumt}
\end{figure}

We now discuss the first (\Fig{cumt}) and the second order cumulant 
(\Fig{2cumt}) in more detail.
These correspond to ensemble average of 
the magnetization and width of the profile, respectively.
We can estimate the second order cumulant of the locked moment profile 
in the absence of magnetic field and the rectangle profile
as $\frac{1}{9}$ and $\frac{S(S+1)}{3S^2}$, respectively. 
The general trend of the first order cumulant looks like more or less
one of either the superparamagnetic or locked moment.
Although, we introduce the present model for describing the intermediate
behavior between superparamagnetic and locked moment, 
the magnetization of intermediate coupling is smaller than 
one of the locked moment.
This is because the Coriolis term decreases the susceptibility, 
as we discussed in Sec.\ref{Mag_Sus}.
We may note, in passing, that the anomalous behavior in the weak coupling,
discussed earlier, should appear when the energy splitting by the magnetic
field comparable to typical energy difference of rotational level, that is,
\eq
2g_{\rm s} B \hbar S \simeq \frac{2\hbar^2 R_{\rm eff}}{2\cJ}
~~\Leftrightarrow~~
g_{\rm s} B \hbar S \simeq \frac{1}{S}
\sqrt{\frac{\hbar^2 S^2 k_{\rm B} T}{2\cJ}}
\simeq 0.14
\en
But it is not seen in the figure \ref{cumt}.

\begin{figure}[htdp]
\centerline{\psfig{figure=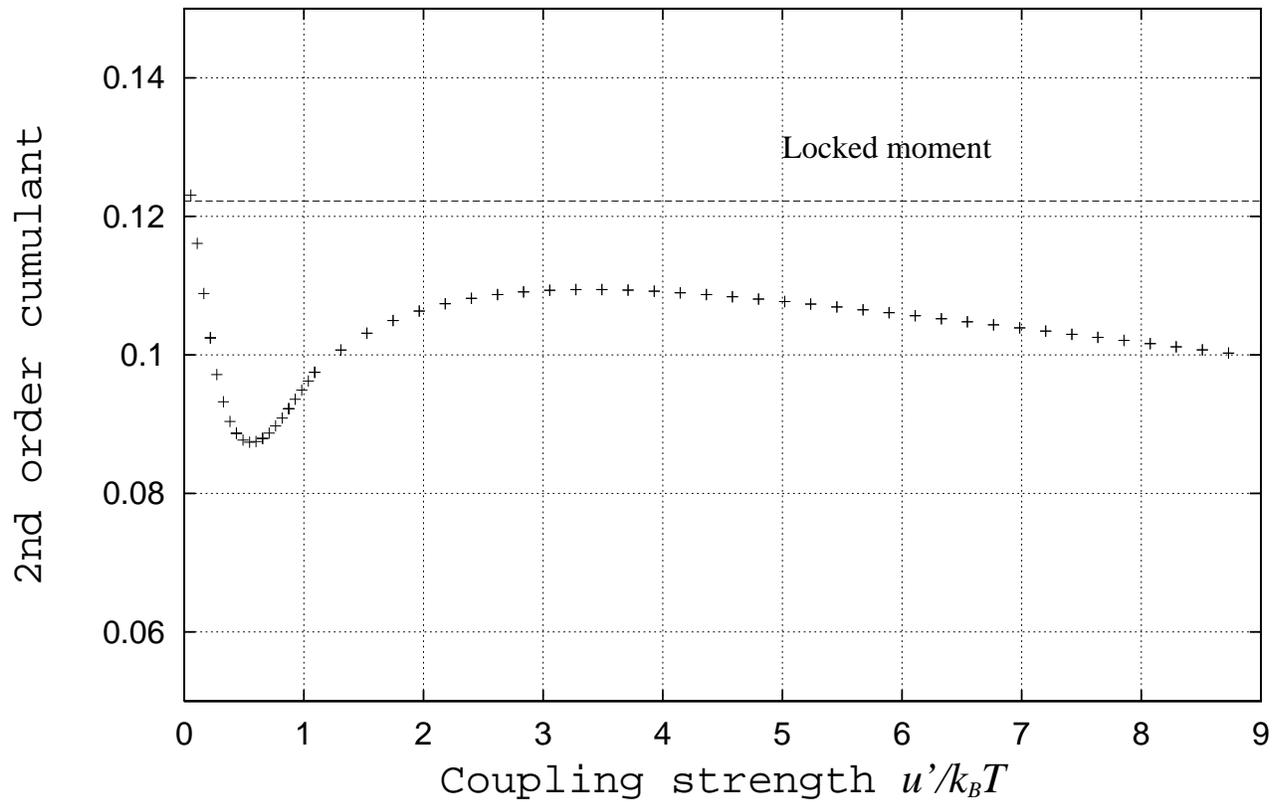,height=12.0cm,angle=-90}}
\caption{The second order cumulant as a function of the ratio of
coupling strength and temperature. The temperature is fixed at 20.}
\label{cumt2}
\end{figure}

The second order cumulant is sensitive to the coupling.
The stronger the coupling becomes, the narrower the profile.
The cut off angular momentum in this calculation is rather small 
than the perturbation.
We calculate cumulant for higher temperature by applying the perturbation
technique.
Figure \ref{cumt2} shows the second order cumulant for 
$I_{\rm max}=80,k_{\rm B} T=20$ as the function of the coupling strength.

The second order cumulant becomes closer to the strong coupling limit $S^2/9$
around the coupling $u'=3.5$. 
It decreases almost linearly in the region of
the coupling beyond $20$ due to the tunneling between different direction
of super-electron spin.
Since the perturbation theory is valid for the magnetic field weaker than 
the coupling strength, the super-spin do not decouple from the cluster.
Therefore, the profile is not the rectangle profile.
We can estimate the second order cumulant of weak coupling limit.
We find the magnetization of each level in this limit from
the diagonal part of \Eq{wm_matrix};
\eq
\bra S_z \ket =\sqrt{2I+1}
 (-1)^{I}\left\langle IM10|IM \right\rangle 
\sqrt{S(S+1)(2S+1)}(-1)^{R-S+1}W(ISIS;R1).
\en
The second order cumulant for weak coupling is estimated as
\eqn
\bra S_z \ket^2_{\rm cumulant} &=&
\frac{S(S+1)(2S+1)}{Z} \sum_{RIM\lm(R)}(2I+1)
| \left\langle IM10|IM \right\rangle |^2 |W(ISIS;R1)|^2 
\exp(-\frac{E_R}{k_{\rm B} T}) \CR
\enn
We treat $R$ as a continuous variable as in the \Subsect{susceptSC} 
and get 
\eqn
\bra S_z \ket^2_{\rm cumulant} &\simeq & \frac{1}{9}S(S+1)
\enn
Actually, in \Fig{cumt2}, the second order cumulant
is not close to the one of the rectangle profile $1/3$, but close
to $\frac{S_{\rm c}(S_{\rm c}+1)}{9 S_{\rm c}^2}\simeq 0.122$.

To make the profile rectangle, the super-spin is decoupled from
the cluster because of the strong magnetic field compare with the coupling;
\eq
g_{\rm s} B \hbar S \gg u'.
\label{sz2_1}
\en
Furthermore, pseudo-crossing do not occur at least for the levels 
having a typical angular momentum. Because, pseudo-crossing 
changes occupation probability of the levels.
\eq
g_{\rm s} B \hbar S < \frac{\hbar^2}{2\cJ}2R_{\rm eff}
\label{sz2_2}
\en
However, in the perturbation for the magnetic field,
the magnetic field is always smaller than the coupling.
Hence, we can not describe the rectangle profile 
by perturbation for the coupling.

To describe the regime around the rectangle profile,
we calculate magnetic field or coupling dependence of $\bra S_z \ket^2$
by perturbation of the coupling under the two assumptions discussed above.
The detailed account of calculation is presented in appendix.
Here we give just the result \Eq{enavSZ2}.
\eqn
\bra S_z \ket^2 &=& \frac{1}{3}S(S+1)
- \frac{8^2}{15 \times 9^2} \left( \frac{u'}{Bg_{\rm s}\hbar} \right)^2 \CR
&&
+\left\{ \frac{4}{9\times 15}\left( \frac{-190+39S(S+1)}{693} \right)
-\frac{4}{15 \times 9^2}S(S+1) \right\}
\left( \frac{u'}{k_{\rm B} T} \right)^2 .
\label{sz2_weak_coupl}
\enn

The higer order cumulant shows the deviation from the gaussian profile.
As a detail of the calculated higher order cumulants from obtained profiles 
are not shown, the result is much smaller than the second order.
When one analizes the experiment in more detail, the higher order cumulants
may bring about important imformation.

\begin{figure}[htdp]
\centerline{\psfig{figure=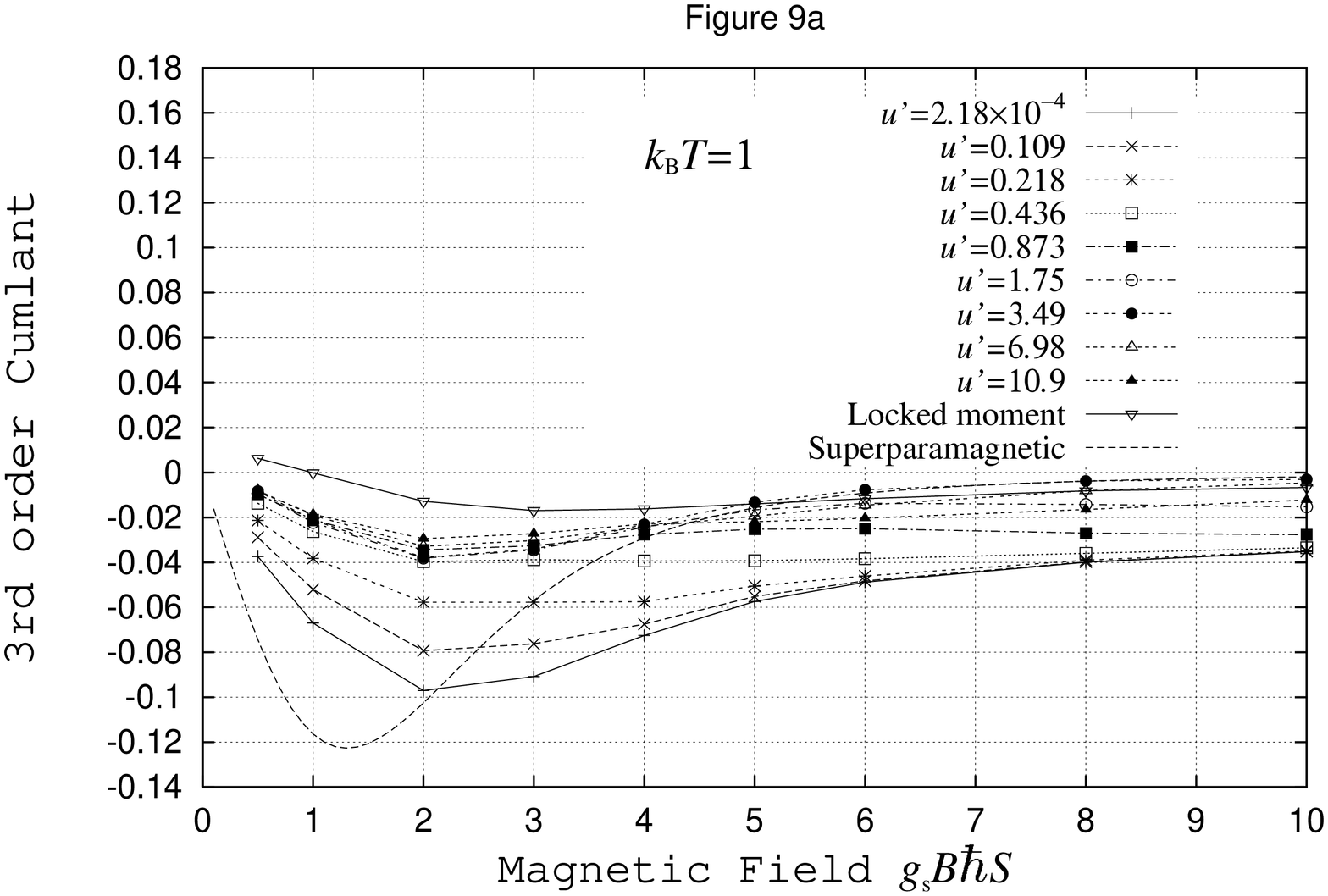,height=10.0cm,angle=-90}}
\centerline{\psfig{figure=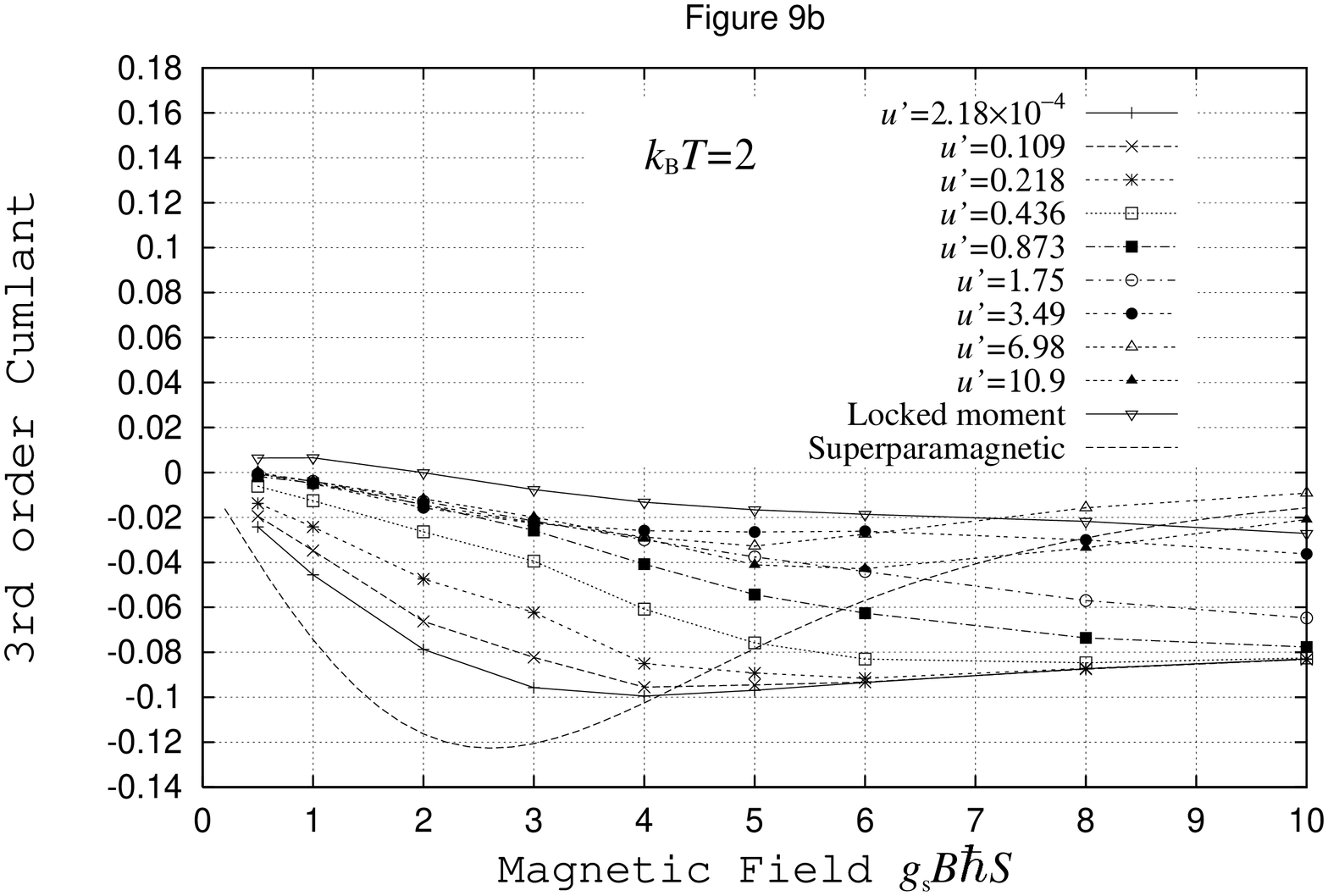,height=10.0cm,angle=-90}}
\caption{The third order cumulant of each profile 
as a function of the ratio of magnetic field. The temperature is fixed at
$k_{\rm B}T=1$ and for $k_{\rm B}T=2$ for Fig. 9a and Fig. 9b respectively.}
\label{3cumt}
\end{figure}

For example, the third order cumulant (\Fig{3cumt})
characterizes asymmetry of the profile.
In the whole range of magnetic field, the third order cumulant is 
smaller than the second order and is smaller as the coupling becomes stronger.
The profiles are always symmetric with respect to $\bra S_z \ket=0$ 
in the absence of magnetic field. 
When the magnetic field is applied, the asymmetry grows on account
of the time reversal symmetry breaking.
The third order cumulant have a peak at $g_{\rm s}B/k_{\rm B} T \simeq 2$.
In the strong magnetic field, the profiles have a narrow peak at
$\frac{\bra S_z \ket}{S}=1$ like \Fig{mag_dep}; 
The profiles are more likely to be symmetric but still not quite.
In other words, the third order cumulant decreases but is not equal to zero.
\section{Conclusions and Discussion}

Using an intermediate coupling model, we have studied the magnetization of
ferromagnetic clusters in a Stern-Gerlach magnet. In this model the
super-electron spin is free but couples to the cluster ions
through an anisotropic potential.
This model is expected to describe the intermediate behavior
between the superparamegnetic and the locked moment.
In evaluating the profiles or the magnetization, we assume that
the variation of the magnetic field in entering the magnet is slow in time,
i.e. adiabatic. Hence, any transition between the quantum states is
suppressed; the occupation probability of each quantum state is determined
in the source area where the magnetic field is absent.

We examined the magnetic susceptibility of the present model
applying perturbation theory. Especially, the magnetic susceptibility
in the strong and in weak coupling limits are discussed analytically.
We expected the intermediate coupling model to approach the locked moment
behavior in the strong coupling limit.
However, there is a crucial difference between the strong coupling 
limit of the present model and the locked moment model.
In the present model the super-electron spin degree of freedom is treated
explicitly and the Coriolis term arises due to the conservation
of total angular momentum, while in the locked moment
model the Coriolis term is neglected.
Consequently, the magnetic susceptibility
is different from, and smaller than the locked moment.
In the weak coupling limit, we expect the susceptibility of the present model 
to become superparamagnetic value. 
But the susceptibility in the present model obtained by the perturbation theory
is not the superparamagnetic value in any range of the coupling.

We find that the irregular magnetic field response noticed in ref \cite{VB96}
also exists in the weak coupling region of the present model.
The magnetization linearly increases before the decoupling;
and is saturates after the decoupling.
Then, once the pseudo-crossing between different $R$ levels occurs,
the magnetization increases again.
Before the decoupling, the susceptibility arises as a
perturbation of the magnetic field. 
But after the decoupling, or once the pseudo-crossing takes place,
the susceptibility becomes nonperturbative.
Instead, we need to diagonalize the total Hamiltonian, which is
difficult for today's computers 
for high temperatures and large cut off angular momentum.
In ref \cite{VB96}, one of the authers (GB) discussed
the susceptibility after the decoupling using uniaxial coupling.
They found that the susceptibility reaches the superparamagnetic value.
Therefore, if we would calculate the susceptibility after the decoupling,
the susceptibility would be superparamagnetic.

Although an anomalous temperature dependence is reported in ref \cite{fe},
the calculated susceptibility is always positive.
We do not reproduce the anomalous temperature dependence.

The `` superparamagnetic peak '' which is seen in the profile 
obtained by the locked moment model 
is not seen in the present calculation even in the strong coupling limit.
As discussed before, the strong coupling limit of our model is
different from the locked moment model due the Coriolis term.
Hence, the effect of Coriolis term destroys the superparamagnetic peak.
If we would be able to calculate in such high temperature
that the Coriolis term can be negligible,
the peak would appear in the deflection profile.


Finally, we summarize our method of analysis 
of the Stern-Gerlach deflection function.
We calculated the cumulant of the profiles up to 3rd order to
characterize the profiles.
We found from figure \ref{cumt}
that the first and second order cumulants are dominant.
In particular, the evaluation of the susceptibility and second order cumulant
by the perturbation technique gave us
the analytical expression of the second order cumulant 
and magnetic susceptibility for high temperature and strong or weak coupling.
One can extract the magnetic moment and the coupling strength by fitting
observed magnetic susceptibility
and the second order cumulant into 
the calculated second order cumulant 
and the magnetic susceptibility.

\section{Acknowledgment}
Part of the calculations were performed with a computer VPP500 at
RIKEN (Research Institute for Physical and Chemical Research, Japan)
and SX4 at RCNP (Research Center of Nuclear Physics).
The authors wish to thank Mr. M.Kaneda and Dr. A.Ansari 
for their helpful advice.
We gratefully acknowledge helpful discussions with Dr. N.Tajima 
especially on the various coupling schemes.
One of the authors (NH) would like to thank A. Bulgac 
for letting him participate to the program ``Atomic Clusters'' (INT-98-2).
This work is finacially supported in part by the Grant-in-Aid 
for Finance Research (09640338).

\appendix
\section{The second order cumulant for the decoupling regime}

In this appendix we derive \Eq{sz2_weak_coupl} under the assumption 
\Eq{sz2_1} and \Eq{sz2_2}.
The ensemble average of the second order cumulant determined through
$\bra S_z \ket^2$ of each level and the energy in the absence of magnetic
field.
Let us start the calculation of second order cumulant 
with evaluating $\bra S_z \ket ^2$.
The assumption \Eq{sz2_1} assures us of applying perturbation theory
of the coupling.
Then we treat $H_{\rm coupl}$ as perturbative and $H_{\rm rot}+H_{\rm mag}$
as nonperturbative.
The degeneracy remains for intrinsic quantum number $k$ 
when we select $\cD^R_{\mu k}(\Omega)\otimes |S\sigma \ket $ 
as unperturbed bases.
According to perturbation theory for degenerate case,
the unperturbed bases are obtained as eigenstates of the Hamiltonian
$H_{\rm coupl}$ of the degenerate space;
\eq
\Psi^{(0)}_{R M \sigma \nu} =\sqrt{\frac{2R+1}{8\pi^2}}
c_\nu^k\cD^R_{\mu k}(\Omega)\otimes |S \sigma \ket,~~~M=\mu+\sigma
\label{unperturbed_base}
\en
The effect of the coupling is included by first- and second-order
perturbation theory.
In the perturbed wave function, there exists mixing with
different rotational states.
Now we should recall the assumption \Eq{sz2_2} in which 
the energy difference between different rotational states
are much larger than the magnetic field.
This justifies that we can neglect the mixing with 
the different rotational states.
Therefore, the perturbed wave function is evaluated as
\eqn
\Psi^{(2)}_{R M \sigma \nu} = \left( 
1 - \frac{1}{2} \sum_{ \sigma \neq \sigma '}
\frac{u'^2|(h_{\rm coupl})^{RM}_{\nu' \sigma',\nu \sigma}|^2}
{\left\{ Bg_{\rm s} \hbar( \sigma -\sigma ')\right\}^2}
\right) \Psi^{(0)}_{R \sigma \nu M} \CR
+  \sum_{ \sigma \neq \sigma '}
\frac{ u'(h_{\rm coupl})^{RM}_{\nu \sigma',\nu \sigma} }
{ Bg_{\rm s} \hbar( \sigma -\sigma ' ) }
\Psi^{(0)}_{R \sigma '  \nu  M}
+ \sum_{ \sigma \neq \sigma '}{\rm (Const.)}
\Psi^{(0)}_{R \sigma ' \nu ' M}.
\label{partWF}
\enn
with
\eqn
u'(h_{\rm coupl})^{RM}_{\nu' \sigma',\nu \sigma}
=
\bra \Psi^{(0)}_{R \sigma ' \nu ' M} | H_{\rm coupl} 
| \Psi^{(0)}_{R \sigma  \nu  M} \ket
=
u'\delta_{\nu,\nu'} \Delta_{\nu R} 
\bra R \mu' 4 m | R \mu \ket
\bra S \sigma 4 m | S \sigma' \ket,
\enn
where  $\Delta_{\nu R}$ represents the eigenvalue of 
$h_{k,k'}=\sum_\kappa \frac{A_\kappa}{A_\pm4}
 \bra R k 4 \kappa | R' k' \ket $ and $u'=A_{\pm 4}
\frac{\bra S || S^4 || S \ket}{\sqrt{2S+1}}$.
Using the perturbed wave function \Eq{partWF}, we calculate $\bra S_z \ket^2$
up to second order for the coupling,
\eqn
\bra S_z \ket^2 &=& \sigma ^2 +\frac{u'^2}{(Bg_{\rm s} \hbar)^2}C_2,\\
&&~~{\rm where}~~C_2=2\sigma \sum_{ \sigma \neq \sigma '}
\frac{|(h_{\rm coupl})^{RM}_{\nu' \sigma',\nu \sigma}
|^2}
{(\sigma ' -\sigma) }.
\label{sz2_expansion}
\enn

Let us move on the evaluation of the energy in the absence of
magnetic field.
We should evaluate the energy which is adiabatically connected with
state in the finite magnetic field.
If the coupling is absent, the unperturbed wavefunction 
\Eq{unperturbed_base} adiabatically connects between
states in the magnetic field and one in the absence of magnetic field.
Hence, we start from unperturbed wavefunction \Eq{unperturbed_base}, 
then introduce the coupling up to the second-order.
According to the perturbation theory, the first order energy shift 
$u'\epsilon ^{(1)} _{\nu R}$ is evaluated the expectation value of 
$H_{\rm coupl}$ for unperturbed state;
\eq
u' \epsilon^{(1)}_{\nu R} =\bra R \mu 4 0 | R \mu \ket 
\bra S \sigma 4 0 | S \sigma \ket u'\Delta_{\nu R}.
\label{energy_shift}
\en
In the second order energy shift, different rotational states mix.
However, as is mentioned above, the assumption \Eq{sz2_2} enables us to
neglect the mixing of different rotational states.
Therefore we neglect second order energy shift.

We are now able to calculate ensemble average of $\bra S_z \ket^2 $
using \Eq{sz2_expansion} and \Eq{energy_shift}.
Expanding not only $\bra S_z \ket$ but also the Boltzmann factor 
up to second order, 
we get the ensemble average of $\bra S_z \ket^2$ as
\eqn
\bra S_z \ket^2_{\rm en.av.} &=& \frac{
\sum_{R \nu \sigma M}
\left(
\sigma^2 + \left(\frac{u'}{Bg_{\rm s}\hbar}\right)^2 C_2 + \cdots~ \right) 
\exp \left( \frac{ -E_R + u' \epsilon^{(1)}_{\nu R} 
+ \cdots }{k_{\rm B} T}
\right)}
{\sum_{R \nu \sigma M}  \exp \left( \frac{ -E_R
+ u' \epsilon^{(1)}_{\nu R} + \cdots }{k_{\rm B} T} \right)
} \CR
&\simeq &
\frac{
\sum_{R \nu \sigma M} \left[
\sigma^2 -\frac{u'}{k_{\rm B} T}\sigma^2 \epsilon^{(1)}_{\nu R} + 
\left\{ \left(\frac{u'}{Bg_{\rm s}\hbar}\right)^2 C_2 + 
\frac{1}{2}\left(\frac{u'}{k_{\rm B} T}\right)^2\sigma^2 {\epsilon^{(1)}_{\nu R}}^2
+\cdots
\right\}
\right] \exp \left( \frac{-E_R}{k_{\rm B} T} \right)
}
{\sum_{R\nu\sigma M}\left\{ 1 - \frac{u'}{k_{\rm B} T}\epsilon^{(1)}_{\nu R} + 
\frac{1}{2}\left(\frac{u'}{k_{\rm B} T}\right)^2{\epsilon^{(1)}_{\nu R}}^2 
+\cdots
\right\} \exp \left( \frac{-E_R}{k_{\rm B} T} \right)
}.
\label{avSZ}
\enn
$\sum_{\nu} \epsilon^{(1)}_{\nu R}= 0$,
$\sum_{\nu} \sigma^2 \epsilon^{(1)}_{\nu R}= 0$,
because Tr $\Delta_{\nu R}=\sum_{\kappa}\bra Rk40|Rk \ket =0$.
Expanding \Eq{avSZ} up to second order of $u'$, we get 
\eq
\bra S_z \ket^2_{\rm en.av.}
= \frac{S(\sigma^2)}{S(1)}
+ \frac{S(C_2)}{S(1)}\left( \frac{u'}{Bg_{\rm s} \hbar}\right)^2
+ \frac{1}{2}\left(
\frac{S(\sigma^2{\epsilon_{\nu R}^{(1)}}^2)}{S(1)}
- \frac{S({\epsilon_{\nu R}^{(1)}}^2)S(\sigma^2)}{S(1)^2}
\right)\left(\frac{u'}{k_{\rm B} T}\right)^2,
\label{sz2enav}
\en
where $S(x)$ is defined as
\eq
S(x)=\sum_{R\nu\sigma M}x\exp\left(-E_R/k_{\rm B} T\right).
\en

Let us estimate $S(x)$'s appearing in \Eq{sz2enav}.
\eqn
S(1) &=& \sum_{R} (2S+1)(2R+1)^2 \exp \left( \frac{-E_R}{k_{\rm B} T} \right).
\label{e0z} \\
S(\sigma^2) &=& \sum_{R} \frac{1}{3}S(S+1)(2S+1)(2R+1)^2
\exp \left(\frac{-E_R}{k_{\rm B} T}\right).
\label{e0s}
\enn
Before calculating the other $S(x)$'s, we evaluate
$\sum_\nu | \Delta_{\nu R} |^2$.
Since the trace of a matrix is constant under the unitary transformation,
we get,
\eqn
\sum_\nu | \Delta_{\nu R} |^2 &=& 
{\rm Tr}(U^{-1} H_{\rm coupl}^R U U^{-1} H_{\rm coupl}^R U ) 
= {\rm Tr}(H_{\rm coupl}^2) \CR
&=& \sum_k ( \frac{14}{5}|\bra R k | 4 0 R k \ket |^2
+|\bra R k | 4 4 R k+4 \ket |^2
+|\bra R k | 4 -4 R k-4 \ket |^2 ) \CR
&=& \frac{8(2R+1)}{15}
\label{sum_epsilon}
\enn
Substituting \Eq{sum_epsilon}, 
we evaluate the other $S(x)$'s
\eqn
\label{e2s}
S({\epsilon_{\nu R}^{(1)}}^2)
&=& \sum_{R}\frac{8(2R+1)}{15}\frac{2R+1}{9}\frac{2S+1}{9}
\exp \left(\frac{-E_R}{k_{\rm B} T} \right).\\
\label{e2ss}
S(\sigma^2 {\epsilon_{\nu R}^{(1)}}^2)
&=& \sum_{R}\frac{8(2R+1)}{15}\frac{2R+1}{9}
\frac{(1+2S)(-190+39S(S+1))}{693}
\exp \left(\frac{-E_R}{k_{\rm B} T} \right).\\
\label{e2z}
S(C_2)
&=& \sum_{R} \frac{8(2R+1)}{15}\frac{2R+1}{9}\frac{-8(2S+1)}{9}
\exp \left(\frac{-E_R}{k_{\rm B} T} \right).
\enn
Finally, putting Eqs. (\ref{e0z}),(\ref{e0s}),
(\ref{e2s}),(\ref{e2ss}),(\ref{e2z})
into \Eq{sz2enav}, we obtain the ensemble average of $\bra S_z \ket^2$ as a
function of the coupling strength.
\eqn
\bra S_z \ket^2 &=& \frac{1}{3}S(S+1)
- \frac{8^2}{15 \times 9^2} \left( \frac{u'}{Bg_{\rm s}\hbar} \right)^2 \CR
&&
+\left\{ \frac{4}{9\times 15}\left( \frac{-190+39S(S+1)}{693} \right)
-\frac{4}{5 \times 9^3}S(S+1) \right\}
\left( \frac{u'}{k_{\rm B} T} \right)^2
\label{enavSZ2}
\enn


\end{document}